\documentclass[range]{ar2e}
\usepackage{graphicx}
\usepackage{epstopdf}
\newcommand{\be}{\begin{equation}}
\newcommand{\ee}{\end{equation}}
\begin{document}

\input epsf.tex    
\input epsf.def   

\input psfig.sty

\jname{Annual Review of Nuclear and Particle Science}
\jyear{2015}
\jvol{}

\title{Topology, magnetic field,  \\ and strongly interacting matter}


\author{Dmitri E. Kharzeev
\affiliation{Department of Physics and Astronomy, \\ Stony Brook University, New York 11794-3800, USA; \\
Department of Physics and RIKEN-BNL Research Center, \\ Brookhaven National Laboratory, Upton, New York 11973-5000, USA}}


\begin{abstract}
Gauge theories with compact symmetry groups possess topologically non-trivial configurations of gauge field. 
This has dramatic implications for the vacuum structure of Quantum Chromo-Dynamics (QCD) and for the behavior of QCD plasma, as well as for condensed matter systems with chiral quasiparticles. I review the current status of the problem with an emphasis on the interplay of chirality with a background magnetic field, and on the observable manifestations of topology in heavy ion collisions, Dirac semimetals, neutron stars, and in the Early Universe.  
\end{abstract}

\maketitle

\section{Introduction}

Quantum Chromodynamics (QCD) possesses a compact gauge group SU(3) that allows for topologically nontrivial solutions for the gluon field. The existence of these solutions changes the vacuum structure  -- 
a superposition of an infinite set of topologically distinct states connected by tunneling instanton  transitions \cite{Belavin:1975fg} becomes the  ``$\theta$-vacuum" of the theory \cite{Callan:1976je,'tHooft:1976fv}.  It is likely that topological effects are linked to confinement and chiral symmetry breaking in QCD, see \cite{Schafer:1996wv} for a review. 

Extended topological configurations of gluon and quark fields in QCD can be effectively probed by a background Abelian magnetic field. The interplay of QCD topology with the topology of zero modes of chiral fermions in a magnetic field leads to a number of surprising novel phenomena that are reviewed below; see the volume \cite{Kharzeev:2013jha} and Refs. \cite{Kharzeev:2013ffa,Shovkovy:2012zn,Andersen:2014xxa,Rebhan:2014rxa} for complementary reviews of topics not covered here in detail. 

Under extreme conditions of high temperature and/or high baryon density, the vacuum of QCD changes its properties, and deconfinement and chiral symmetry restoration take place. This deconfinement transition is accompanied by the rapid change in the rate and nature of topological transitions connecting different topological sectors, as discussed in section \ref{above_bar}. 
The heavy ion program reviewed recently in \cite{Shuryak:2014zxa} opens a possibility to study these phenomena in experiment.

 Moreover, since the colliding ions create strong magnetic fields ${\cal O}(10\ m_{\pi}^2)$ \cite{Kharzeev:2007jp} (see \cite{Tuchin:2013ie} for review), the interplay of QCD topology with an Abelian magnetic background can be studied experimentally,  as discussed in section \ref{cme_hic}. The existence of strong magnetic fields in neutron stars allows to study novel topological phenomena in cold and dense QCD matter that are discussed in section \ref{neutron}. The opposite limit of high temperature and small baryon density is reached in the Early Universe, see section \ref{cme_prim}.

The recent discovery of Dirac semimetals -- see \cite{Ong} for a review --- opens a fascinating possibility to study the topological phenomena involving 3D chiral fermions in condensed matter experiments. In section \ref{cme_condsec} we describe the recent observation of the chiral magnetic effect in ZrTe$_5$ \cite{Li:2014bha}.

\section{Above the barriers: Chern-Simons number diffusion}\label{above_bar}

\subsection{Topology across the QCD phase transition}\label{top_across}

The existence of the degenerate vacuum states with different Chern-Simons numbers induces a  P- and CP-odd ``$\theta$-term" in the lagrangean of $SU(N)$ gauge theory:
\be
{\cal L}_{\rm QCD} =  -{1 \over 4} G^{\mu\nu}_{\alpha}(x)G_{\alpha \mu\nu}(x)  - \theta\ q(x) ,
\ee
where 
\be 
q(x) = {1 \over 32 \pi^2}\  g^2 G^{\mu\nu}_{\alpha}(x) \tilde{G}_{\alpha \mu\nu}(x)
\ee
is the density of topological Chern-Pontryagin charge, and $\theta$ angle is a new parameter of the theory \cite{Callan:1976je}. The absence of observed global P and CP violation in QCD indicates that $\theta$ is very close to zero -- however the requirement of periodicity with respect to this angular variable puts important constraints on the QCD dynamics.

At zero temperature, reconciling large $N$ (number of colors) scaling with periodicity dictates \cite{Witten:1998uka} that the vacuum energy $E(\theta)$ is a multi-valued function of $\theta/N$:
\be\label{multiv}
E(\theta) = N^2\ {\rm min}_k\ \bar{E}\left(\frac{\theta + 2 \pi k}{N}\right), 
\ee 
where $k$ is an integer. As $\theta$ varies adiabatically, the ground state moves from one branch of the potential Eq. (\ref{multiv}) to the other; note that this implies the existence of unstable vacuum states.

\begin{figure}
\begin{center}
\includegraphics[width=12cm]{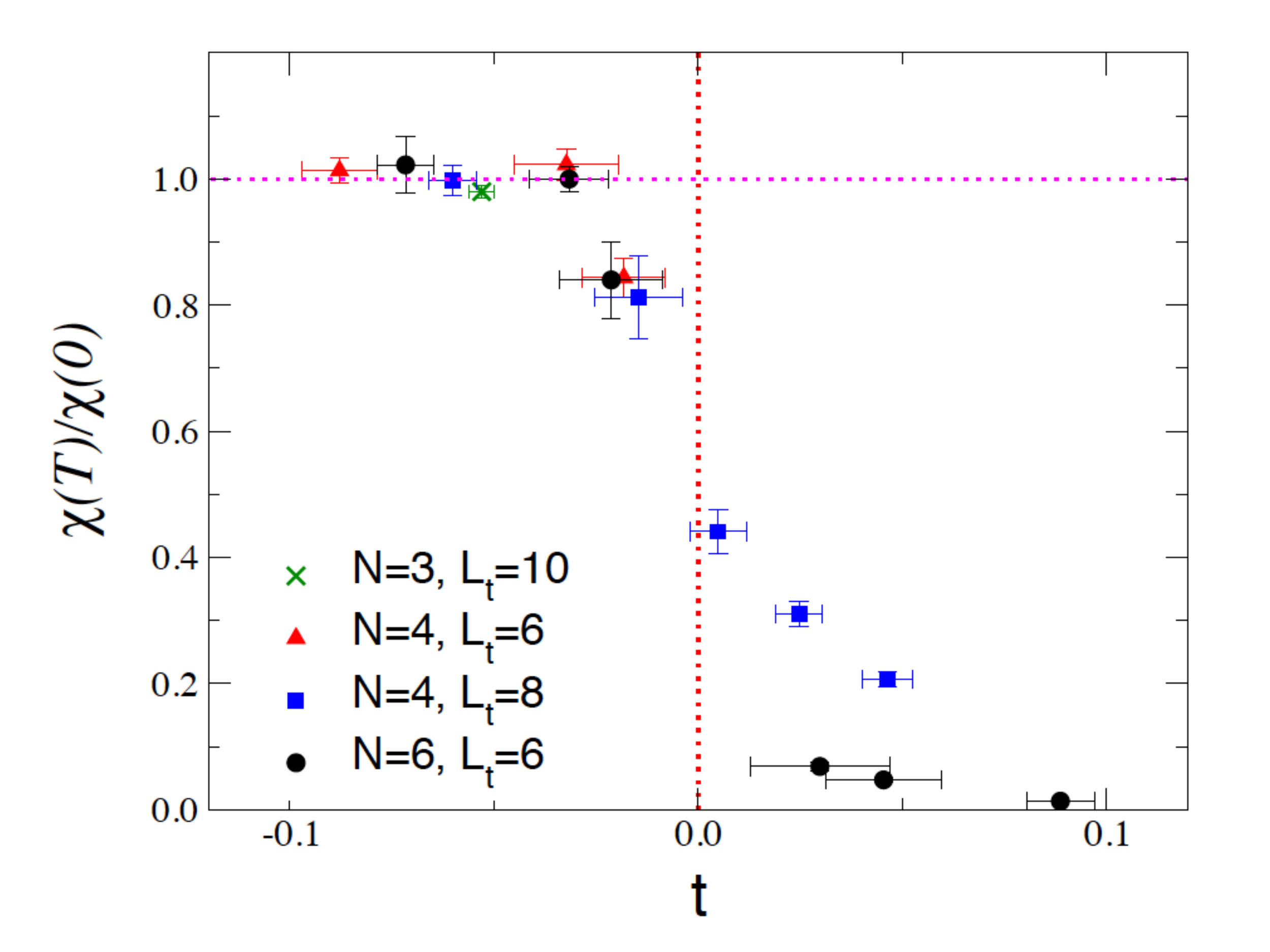}
\end{center}
\caption{The temperature dependence $t = T/T_c -1$ of the ratio $\chi(T)/\chi(0)$ of topological susceptibilities at non-zero and zero temperature $T$ in SU(N) gauge theory; from \cite{Bonati:2013tt,Bonati:2013dza}.}
\label{top_sus}
\end{figure}

At small $\theta$, we can  write the free energy at a finite temperature $T$ as
\be
F(\theta, T) - F(0, T) = \frac{1}{2} \chi(T)\ \theta^2\ f(\theta, T),
\ee
where $\chi(T)$ is the topological susceptibility. Expanding around $\theta = 0$, we can further write the function $f(\theta, T)$ as
\be
f(\theta, T) = 1 + b_2\ \theta^2 + ... ;
\ee
only even powers of $\theta$ are allowed by the P- and CP-invariance of the ground state. The temperature dependences of topological susceptibility and of the moment $b_2$ encode important information about the changes in the topological structure of the vacuum, and can be studied in lattice QCD simulations. The behavior of topological susceptibility as a function of temperature has been computed recently \cite{Bonati:2013tt,Bonati:2013dza} in $SU(N)$ gauge theory; the results are shown in Fig. \ref{top_sus}. One can see that the topological susceptibility rapidly decreases around the deconfinement transition, and this dependence sharpens at large $N$ in accord with the scenario proposed in  \cite{Kharzeev:1998kz}. The rapid decrease of topological susceptibility  implies the existence of metastable P- and CP-odd vacuum states near $T_c$ \cite{Kharzeev:1998kz}. 

\begin{figure}
\begin{center}
\includegraphics[width=12cm]{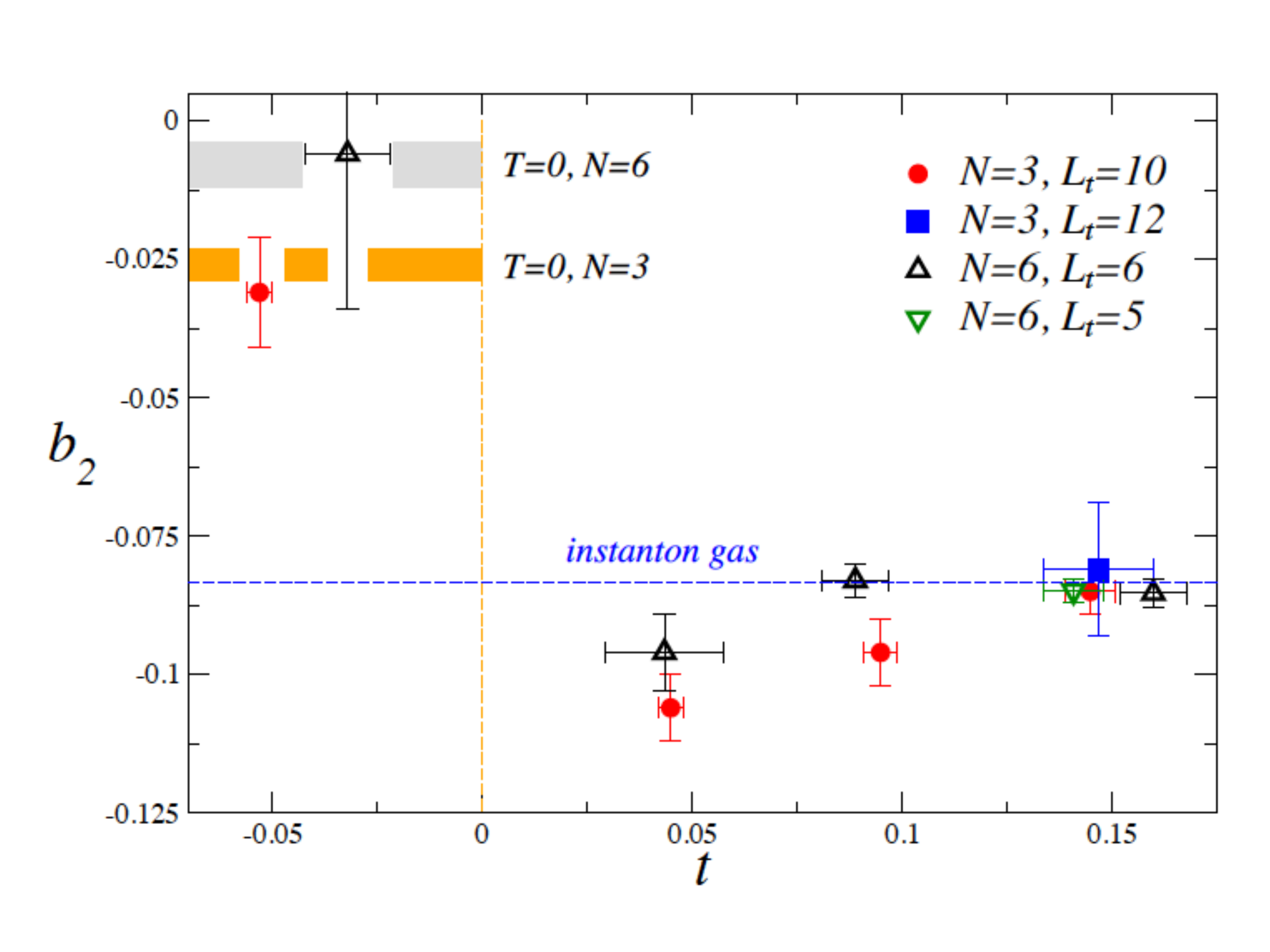}
\end{center}
\caption{The temperature dependence $t = T/T_c -1$ of the $b_2$ moment of the topological charge distribution in SU(N) gauge theory for N = 3 and N = 6; from \cite{Bonati:2013tt}.}
\label{b2}
\end{figure}

To gain an additional insight, let us examine the lattice results \cite{Bonati:2013tt,Bonati:2013dza} on the temperature dependence of the moment $b_2$ presented in Fig. \ref{b2}. At high temperatures, color screening suppresses instantons with size $\rho \gg 1/T$ \cite{Gross:1980br, Shuryak:1978yk, Pisarski:1980md}, so the dilute instanton gas picture should be appropriate, see \cite{Schafer:1996wv} for review. In this picture, the $\theta$ dependence of the free energy is given by
\be
F(\theta, T) - F(0, T) \simeq \chi(T)\ (1 - \cos \theta),
\ee
which yields a definite prediction for the moment $b_2$. As can be seen from Fig. \ref{b2}, the dilute instanton gas approximation is indeed appropriate for high temperatures in the deconfined phase; this regime is approached faster at large $N$. However at low $T$ and around the deconfinement transition the topological structure of QCD matter is not captured by the instanton gas picture -- this implies that the effects of confinement on topological fluctuations are substantial and have to be understood.

\subsection{Sphaleron rate in hot QCD plasma: weak coupling}

At finite 
 temperature, the transitions between the vacuum states with different topological Chern-Simons numbers can be induced by a classical thermal activation process, so-called "sphaleron" \cite{Klinkhamer:1984di}. In electroweak theory sphaleron transitions cause the baryon number violation and may be responsible for at least a part of the observed baryon asymmetry in the Universe \cite{Kuzmin:1985mm}; for a review, see \cite{Rubakov:1996vz}. In QCD plasma, sphalerons  are abundant \cite{McLerran:1990de} and induce the quark chirality non-conservation. 
 
 Unlike for the tunneling instanton transitions, the rate of the sphaleron transitions $\Gamma$ is not exponentially suppressed at weak coupling $g$, and in Yang-Mills theory with $N$ colors is proportional to \cite{Arnold:1996dy,Huet:1996sh,Bodeker:1998hm} 
\be\label{weak}
 \Gamma = const \times (g^2 N)^5 \ln(1/g^2 N) \ T^4 ,
\ee
 with a numerically large pre-factor \cite{Moore:1997sn,Moore:2010jd}. Sphalerons describe a random walk in the topological Chern-Simons number $N_{CS}$:
  \be
  \Gamma = \lim_{T \to \infty} \frac{(N_{CS}(T) - N_{CS}(0))^2}{V T} ,
  \ee
  so that in a volume $V$ and after a (sufficiently long) time period $T$ we get the topological number $\langle N^2_{CS} \rangle =  \Gamma\ V\ T$. 
  
  The expression Eq.(\ref{weak}) can be qualitatively understood in the following way:  since the sphaleron at the peak of the barrier separating Chern-Simons sectors is a purely magnetic field configuration, the factor $(g^2 T)^3$ is the inverse magnetic screening length that determines the characteristic inverse volume of the sphaleron, and $g^4 T \log(1/g)$ is the typical inverse time scale of the process.

  The existing weak coupling computations are based on semiclassical field theory approaches, and do not provide a reliable result for the values of strong coupling $\alpha_s \sim 1/2$ relevant for quark gluon plasma produced in heavy ion collisions. A ``desperate extrapolation" \cite{Moore:2010jd} of the weak coupling results yields an estimate 
 \be\label{pert_rate}
 \Gamma \simeq 30\ \alpha_s^4\ T^4 .
 \ee  

\subsection{Chern-Simons diffusion in non-Abelian plasma at strong coupling: holography} 

At strong coupling, a valuable insight into the dynamics of non-Abelian conformal theories is offered by the holographic AdS/CFT correspondence \cite{AdS-CFT,AdS-CFT1,AdS-CFT2}. The holographic correspondence establishes duality between a strongly coupled conformal field theory in the boundary Minkowski space-time and classical supergravity in the AdS$_5$ bulk space.  The emergence of a black hole in the AdS$_5$ space corresponds to the formation of plasma with temperature equal to the Hawking temperature of the black hole. 

This approach allows to evaluate the Chern-Simons diffusion rate in conformal 
 ${\cal N} =4$ maximally super-symmetric Yang-Mills theory (${\cal N}=4$ SYM) by considering the bulk propagation of axions (excited by the topological charge operator on the boundary) in the background of the black hole \cite{Son:2002sd}. The absorption of the axions by the black hole with a cross section equal to the area of the black hole delivers the following result \cite{Son:2002sd}:
  \be\label{sphaleron}
 \Gamma = \frac{(g^2 N)^2}{256\ \pi^3}\ T^4 ,
 \ee
showing that topological transitions become more frequent at strong coupling, even though the dependence on the coupling is weaker than suggested by Eq.\ref{weak}. This means that Chern-Simons diffusion is not necessarily associated with semi-classical field configurations that are well defined only at small coupling and are usually washed out by quantum effects at strong coupling. Note that the large $N$ behavior is the same in the weak and strong coupling limits ($\sim N^0$). 

Of course, QCD is not a conformal theory, and the quark-gluon plasma at temperatures $T \leq 2\ T_c$ significantly deviates from the scale-invariant behavior, as signaled by the expectation value of the trace of the energy-momentum tensor $\langle \theta_\mu^\mu \rangle_T = \epsilon - 3 P$ measured on the lattice \cite{Bazavov:2014pvz} -- the energy density $\epsilon$ and pressure $P$ of relativistic conformal plasma are related by $\epsilon = 3 P$. It is thus interesting to establish the effect of this deviation from conformality on the Chern-Simons diffusion rate. 

This question has been addressed \cite{Gursoy:2012bt} recently within the ``improved holographic QCD" \cite{Gursoy:2010fj} - a bottom-up approach describing the running QCD coupling and confinement by a properly chosen dilaton potential in the bulk. The resulting Chern-Simons diffusion rate \cite{Gursoy:2012bt} is significantly larger than that given by Eq.(\ref{sphaleron}): at $\alpha_s = 0.5$, the authors find 
$\Gamma(T_c) \geq 1.64\ T_c^4$ at the transition temperature. This is about 36 times larger rate than given by Eq.(\ref{sphaleron}) at $\alpha_s = 0.5$, and is about the same (within a factor of two) as the weak coupling extrapolation  Eq.(\ref{pert_rate}). 

It is important to identify configurations responsible for Chern-Simons diffusion at strong coupling. 
An explicit example of a topological solution in semiclassical $AdS_5 \times S_5$ supergravity (dual to the large $N$, strongly coupled ${\cal N}=4$ SYM) is provided by the D-instanton \cite{Gibbons:1995vg,Green:1997tv} that is obtained from solitonic D-brane solutions by wrapping them around an appropriate compact manifold. D-instanton can be viewed as an Einstein-Rosen wormhole connecting two asymptotically Euclidean regions of space-time, with the Ramond-Ramond (R$\otimes$R) charge flowing down the throat of the wormhole \cite{Gibbons:1995vg}. It describes a process of violation of the conservation of a global charge in the boundary theory \cite{Gibbons:1995vg}.  D-instantons have been proposed \cite{Kharzeev:2009pa} as a source of multiparticle production in high energy collisions in strongly coupled ${\cal N}$=4 SYM, extending on the earlier ideas of   \cite{Kharzeev:1999vh,Kharzeev:2000ef,Shuryak:2000df,Nowak:2000de} based on the weak coupling approach.

\subsection{QCD phase diagram in the $(T, \mu_5)$ plane}

Because the chiral anomaly relates the net chirality of fermions to the topology of gauge fields, Chern-Simons diffusion generates the quark chirality imbalance. This imbalance can be quantified by 
the chiral chemical potential $\mu_5 = (\mu_R - \mu_L)/2$ that describes an asymmetry between the Fermi energies of left- and right-handed quarks. The phase diagram of QCD in the $(T, \mu_5)$ plane can thus shed light on the connection between confinement and topology. 

Moreover, since the chiral chemical potential does not lead to the fermion determinant sign problem \cite{Fukushima:2008xe}, this phase diagram can be studied in the first principle simulations on the lattice. Note however that because of the chiral anomaly, the chiral chemical potential does not correspond to a conserved charge -- the decay of chirality into gauge fields with non-trivial topology induces the so-called ``chiral magnetic instability", see discussion in section \ref{neutron}.

\begin{figure}
\begin{center}
\includegraphics[width=15cm]{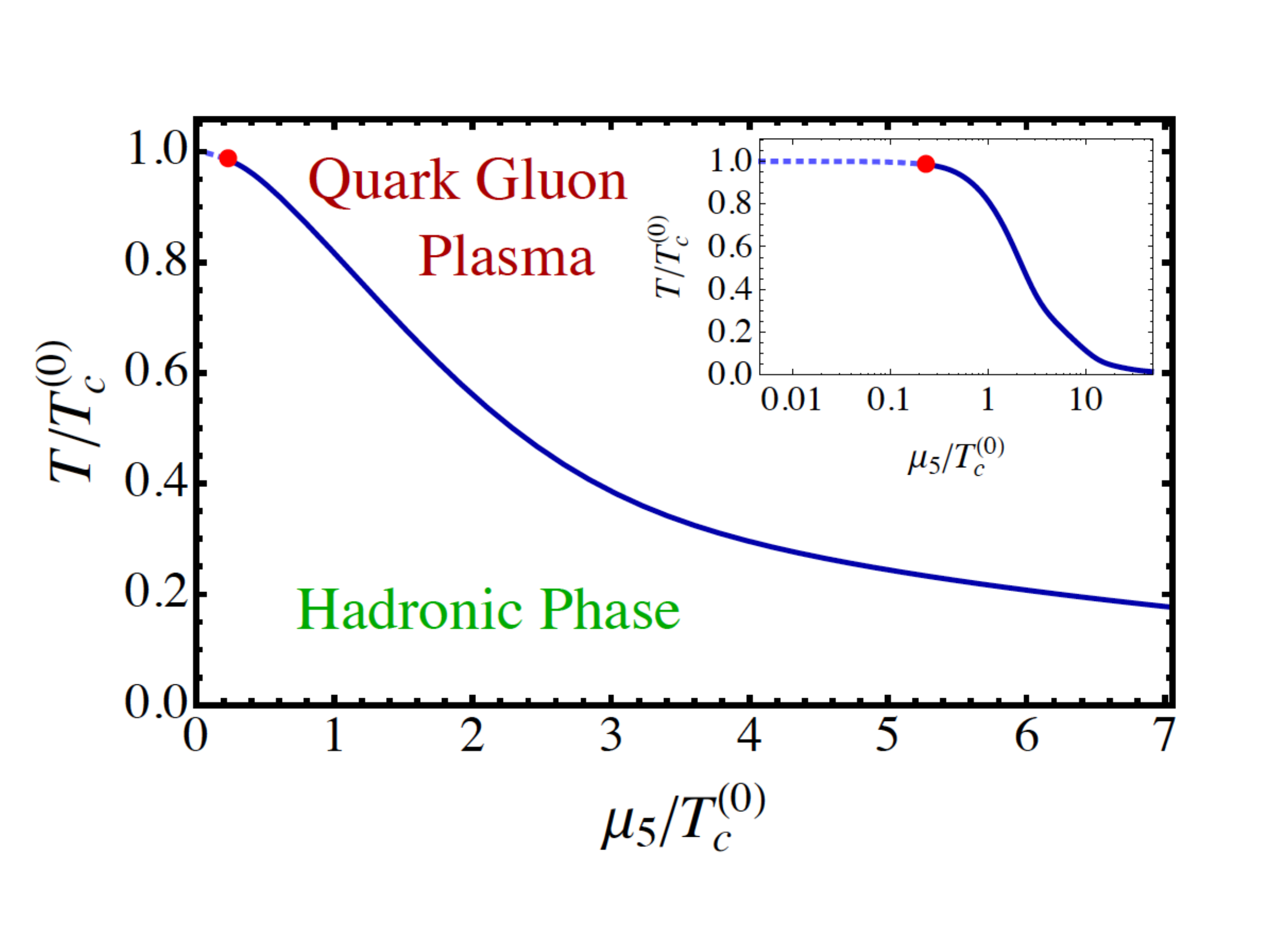}
\end{center}
\vspace{-1cm}
\caption{The conjectured phase diagram of QCD in the temperature -- chiral chemical potential plane;  from \cite{Chernodub:2011fr}.}
\label{phasediag5}
\end{figure}

It has been argued in \cite{Fukushima:2010fe,Chernodub:2011fr,Gatto:2011wc} that at finite $\mu_5$ the critical temperature decreases, and at some value of $\mu_5$ the phase transition becomes first order via the second order end-point, in analogy to the behavior expected \cite{Stephanov:1998dy} in the $(T,\mu)$ plane, see Fig. \ref{phasediag5}. The first lattice studies at finite $\mu_5$ have recently been performed in $SU(2)$ theory \cite{Braguta:2014ira} with quark masses corresponding to the pion mass in the range $m_{\pi} \sim 300 \div 500$ MeV. Contrary to the model expectations \cite{Fukushima:2010fe,Chernodub:2011fr,Gatto:2011wc}, the lattice results indicate that the critical temperature for the deconfinement and chiral symmetry restoration {\it increases} at finite $\mu_5$. It remains to be seen if this conclusion persists at physical quark masses -- if it does, the result would imply a  mechanism of interplay between  confinement and topology not captured by current models. 

\section{QCD in an external magnetic field}\label{qcd_magf}

\subsection{QCD phase diagram in magnetic field}

An external magnetic field provides a unique way to probe QCD topology; it is thus instructive to study the phase diagram of QCD in a magnetic background. One of the key phenomena affecting chiral fermions in magnetic field is the ``magnetic catalysis" \cite{Gusynin:1994re}, see \cite{Shovkovy:2012zn} for a recent review. This term refers to the enhancement of dynamical symmetry breaking induced by an external magnetic field. In a strong magnetic field, this phenomenon can be understood in terms of the effective $D \to D-2$ dimensional reduction. In particular, magnetic field promotes the formation of scalar chiral condensate -- since the spins of fermions and antifermions on the lowest Landau level point in opposite directions, the formation of a spin-zero condensate  is favored. Note however that the enhancement of symmetry breaking in magnetic field does not necessarily involve Landau levels -- for example, the {\it in-plane} magnetic field in graphene can induce excitonic instability \cite{Aleiner:2007va}.

In a relatively weak magnetic field $m_\pi^2 < eB < 4 \pi F_\pi^2$ ($F_\pi$ is the pion decay constant), the correction to the chiral condensate can be evaluated using the chiral perturbation theory \cite{Shushpanov:1997sf}
\be\label{cond_mag}
\Sigma(B) = \Sigma(0) \ \left(1 + \frac{eB\ \ln 2}{(4 \pi F_\pi)^2} + {\cal O}\left(\frac{(eB)^2}{F_\pi^4}\right) \right) ;
\ee
this result demonstrates an increase of the condensate in magnetic field. Note that the dependence on the field in Eq. (\ref{cond_mag}) is linear, and not quadratic as can be naively expected from power counting. This is because the pion loop diagrams  describing the vacuum energy density in magnetic field yield logarithms involving the infrared cutoff that is usually provided by the pion mass. The chiral condensate can be obtained from the energy density by differentiating w.r.t. the quark mass, or, using the GMOR relation, w.r.t. the pion mass squared. As a result, the chiral condensate becomes inversely proportional to the infrared cutoff. However in an external magnetic field, the appropriate physical infrared cutoff becomes $eB > m_\pi^2$  -- so the naive quadratic dependence on $eB$ gets replaced by the linear one \cite{Shushpanov:1997sf}. In the absence of Goldstone modes, the dependence of the condensate on $eB$ is quadratic \cite{Klevansky:1989vi}, as naively expected. 
This discussion illustrates why an external magnetic field  has a big effect on QCD interactions -- it provides a physical (and anisotropic) infrared cutoff, and QCD possesses a large sensitivity to the infrared regularization.

\begin{figure}
\begin{center}
\includegraphics[width=15cm]{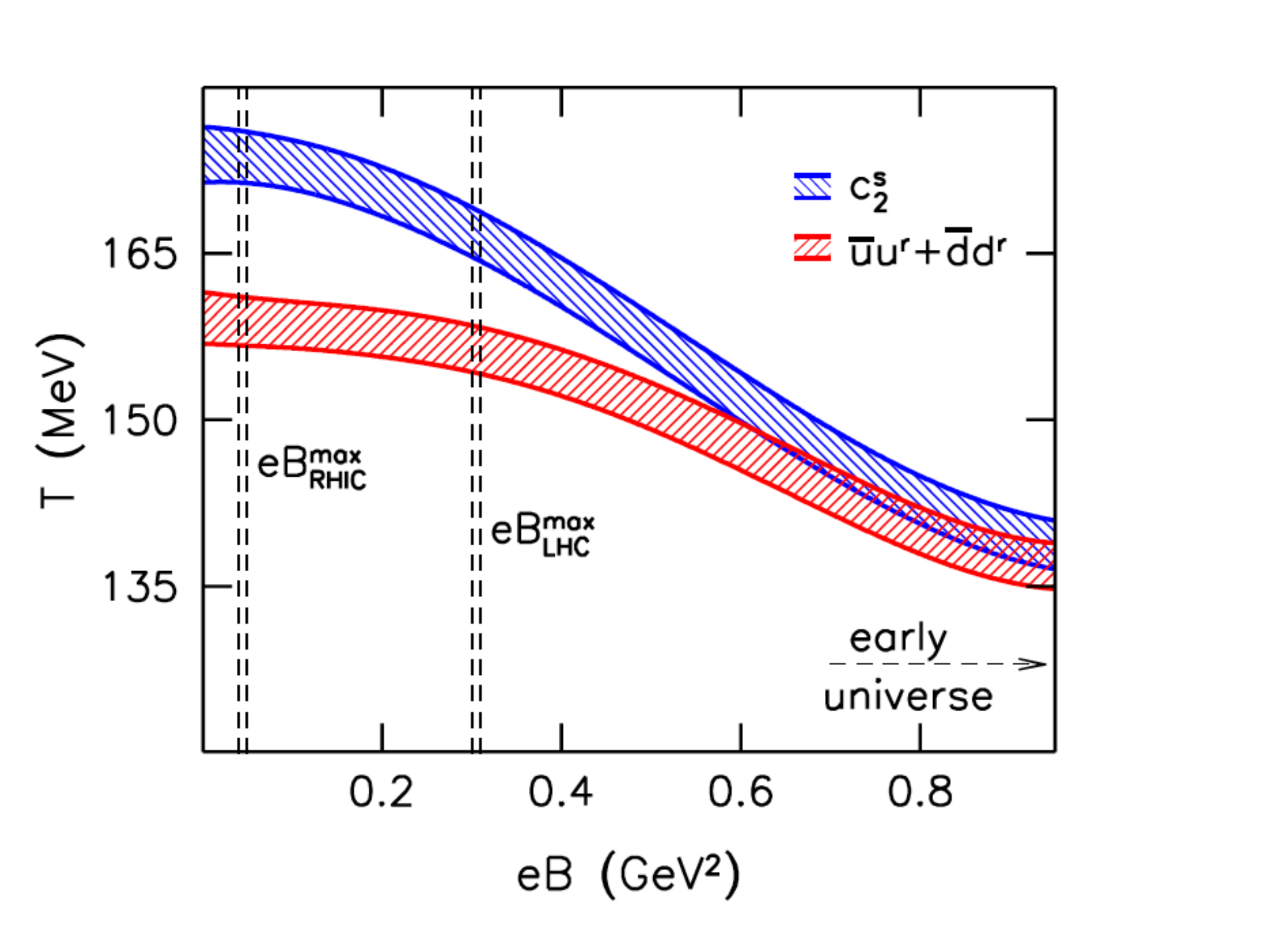}
\end{center}
\caption{The phase diagram of QCD in the temperature -- magnetic field $(T, eB)$ plane determined from $N_f = 2+1$ lattice QCD;  from \cite{Bali:2011qj}.}
\label{phase_diag_mag}
\end{figure}

The lattice studies of QCD in magnetic field indeed indicate \cite{Buividovich:2008wf} the expected increase of the chiral condensate in magnetic field consistent with Eq.(\ref{cond_mag}). On the basis of magnetic catalysis, it has also been widely expected that the restoration of chiral symmetry in QCD will be delayed to higher temperatures in an external magnetic field. 
However the first principle lattice QCD $\times$ QED results indicate the opposite trend, see Fig. \ref{phase_diag_mag} from \cite{Bali:2011qj}. The lattice results show that, surprisingly,  a phenomenon opposite to magnetic catalysis takes place in QCD matter -- this has been called the ``inverse magnetic catalysis"; the possible origins of this effect have been discussed in \cite{Bruckmann:2013oba,Andersen:2014xxa}. 

The properties of the QCD equation of state in a background magnetic field have recently been studied in \cite{Bali:2014kia}, with intriguing results. In particular, while QCD matter appears weakly diamagnetic at low temperatures (consistent with the dominance of pions), around and above the transition temperature the response is paramagnetic \cite{Bali:2014kia}. This means that the QCD plasma is susceptible to magnetization, e.g. in the form of helical magnetic configurations triggered by the chiral magnetic instability, see Section \ref{neutron}. The studies of QCD matter in magnetic field are discussed in much more detail in recent reviews \cite{Fraga:2012rr,Kharzeev:2013jha,Andersen:2014xxa,Rebhan:2014rxa}.

\subsection{Instantons in magnetic field}\label{instan_mag}

Instantons \cite{Belavin:1975fg} are configurations of the gluon field and thus do not interact directly with an external Abelian magnetic field.  
However quarks carry both electric and color charge and couple to both magnetic and instanton fields. A magnetic field introduces Landau levels in which the fermion zero modes have a definite spin projection on the magnetic field. In the instanton background the fermion spectrum also has zero modes, with chiralities determined locally by the local topological charge.  

The combination of background fields of the instanton and an Abelian magnetic field thus leads to a surprisingly intricate and rich structure in the Dirac spectrum studied in \cite{Basar:2011by}. The result is driven by the competition of two effects: the spin projection produced by a magnetic field and the chirality projection produced by an instanton field. Of particular interest is the emergence of electric dipole moment in a magnetic field that signals the violation of P and CP symmetries localized around the instanton \cite{Basar:2011by}. The easiest way to understand this phenomenon is to use the chiral representation of the quark spinors and to write down the electric dipole moment as
\be\label{eldipes}
\sigma_3^E = - {\bar{\psi}}_L \sigma_3 \psi_L + {\bar{\psi}}_R \sigma_3 \psi_R ;
\ee
magnetic field is directed along $x_3$, and $\sigma$ is the spin operator. As can be seen from Eq.\ref{eldipes}, in the absence of chirality asymmetry the contributions from left- and right-handed quarks cancel each the electric dipole moment is equal to zero. However in the presence of the instanton the left-right symmetry is lost, and this leads to the emergence of the electric dipole moment. In the limit of strong magnetic field $eB \gg \rho^{-1}$ ($\rho$ is the instanton size), the Dirac spectrum can be evaluated exactly \cite{Basar:2011by}.

\begin{figure}
\begin{center}
\includegraphics[width=15cm]{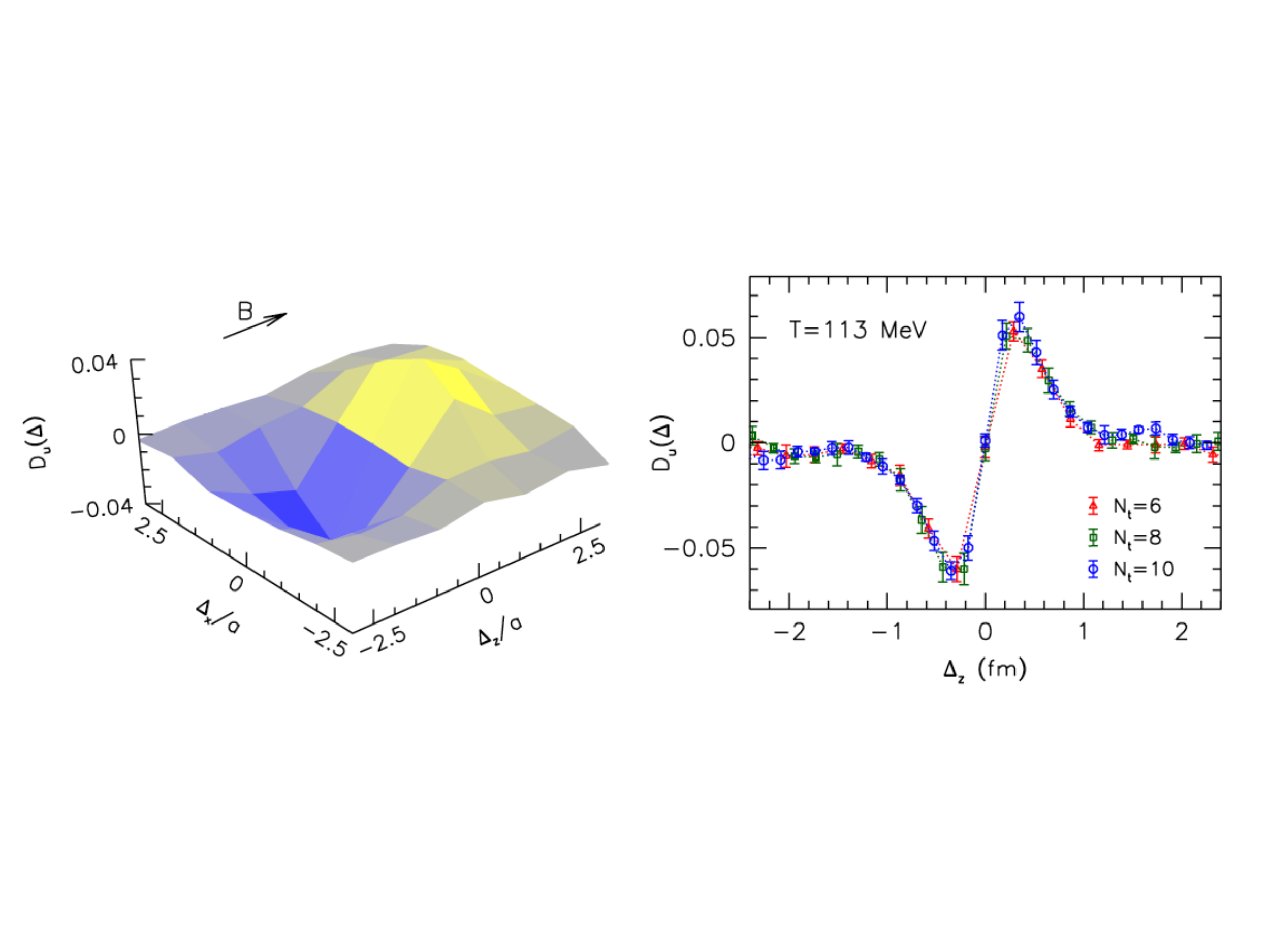}
\end{center}
\vspace{-3cm}
\caption{The electric dipole structure associated with the topological charge in QCD matter (left panel); 
the spatial correlation of the electric and topological charges along the direction of magnetic field at temperature T = 113 MeV;  from \cite{Bali:2014vja}.}
\label{lattice_CME}
\end{figure}

Lattice calculations in QCD $\times$ QED confirm the emegence of electric dipole moment around the topological fluctuations of gluon field \cite{Buividovich:2009wi,Buividovich:2009my,Buividovich:2010tn,Abramczyk:2009gb,Bali:2014vja}. Fig. \ref{lattice_CME} shows the electric dipole emerging around a topological fluctuation in QCD matter at temperature $T = 113$ MeV \cite{Bali:2014vja}.

\subsection{Chern-Simons diffusion in magnetic field}

The Chern-Simons diffusion rate for strongly coupled ${\cal N}=4$ plasma in the presence of an external magnetic field was computed via holography in Ref. \cite{Basar:2012gh}. The dual gravity is characterized by a full solution of asymptotically $AdS_5$, five-dimensional Einstein-Maxwell system with a constant magnetic flux.  This solution takes into account all of the back-reaction of the magnetic field on the metric and allows one to work with arbitrarily strong magnetic fields; it was introduced in \cite{D'Hoker:2009mm}, see also \cite{D'Hoker:2012ih}.

In a weak magnetic field, to leading order in $B^2$ one gets \cite{Basar:2012gh}:
\be
\Gamma_{CS}= \frac{(g^2N)^2}{256 \pi^3}T^4\left(1+ \frac{1}{6\pi^4}\,\frac{B^2}{T^4}+{\cal O} \left(\frac{B^4}{T^8}\right)\right) ;
\ee 
note that at zero magnetic field this expression coincides with Eq.\ref{sphaleron}. The pre-factor of $B^2/T^4$ is numerically small $\sim 2\ 10^{-3}$, which suggests that the effect of magnetic fields on the sphaleron rate in heavy ion collisions may be neglected.

In the strong magnetic field limit the diffusion rate is 
\be
\Gamma(B,T)=\frac{(g^2N)^2}{384 \sqrt{3} \pi^5} B\, T^2\quad,\quad B>>T^2 .
\label{strongbrate}
\ee
The result Eq.\ref{strongbrate} can be understood as a consequence of dimensional reduction in strong magnetic field. In holographic approach, this is manifested through the emergence of a (2+1)-dimensional BTZ black hole in the bulk \cite{Basar:2012gh}. It is interesting that this dimensional reduction applies even to the fields that are not charged under the external magnetic field -- this is a consequence of strong coupling. 

\section{The chiral magnetic effect (CME)}

The term Chiral Magnetic Effect \cite{Kharzeev:2007jp,Fukushima:2008xe} refers to 
 the generation of electric current induced by the 
chirality imbalance in the presence of an external magnetic field:
\be\label{cme_current}
{\vec J} = \frac{e^2}{2 \pi^2}\ \mu_5\ {\vec B},
\ee
where $\mu_5$ is the chiral chemical potential. In QCD matter subjected to an external magnetic field or rotation, topological fluctuations thus induce fluctuations of the electric dipole moment \cite{Kharzeev:2004ey,Kharzeev:2007tn,Kharzeev:2007jp,Fukushima:2008xe}; we have already encountered a phenomenon of this type in section \ref{instan_mag} addressing QCD instanton in a magnetic background. 

The physics of CME and related phenomena has been recently reviewed in \cite{Kharzeev:2013ffa} which also contains an extensive set of references. In this paper we will thus limit ourselves only to a brief summary followed by the discussion of recent CME-related developments in  heavy ion physics (section \ref{cme_hic}), condensed matter physics (where we describe the first observation of CME in section \ref{cme_condsec}), physics of neutron stars (section \ref{neutron}) and the Early Universe (section \ref{cme_prim}). 

The CME is a macroscopic quantum effect - it is a manifestation of the chiral anomaly creating a collective motion in Dirac sea. 
Because the chirality imbalance is related to the global topology of gauge fields, the CME current is topologically protected \cite{Kharzeev:2009fn} and thus 
{\it non-dissipative} even in the presence of strong interactions. As a result, the CME and related quantum phenomena affect the hydrodynamical 
and transport behavior of systems possessing chiral fermions, from the quark-gluon plasma to Dirac semimetals.

The persistence of CME at strong coupling and small frequencies makes the hydrodynamical description of the effect possible, and indeed it arises naturally within the relativistic hydrodynamics as shown by Son and Surowka \cite{Son:2009tf}.  The quantum anomalies in general have been found to modify hydrodynamics in a significant way, see 
\cite{Zakharov:2012vv} for a review.  The principle of ``no entropy production from anomalous terms" can be used to constrain the relativistic conformal hydrodynamics at second order  in the derivative expansion, where it allows to compute analytically 13 out of 18 anomalous transport coefficients \cite{Kharzeev:2011ds}. 

Anomalous hydrodynamics has been found to possess a novel gapless collective excitation -- the ``chiral magnetic wave" \cite{Kharzeev:2010gd}, see also \cite{Newman:2005hd}. It is analogous to sound, but in strong magnetic field propagates along the direction of the field with the velocity of light \cite{Kharzeev:2010gd}. The chiral magnetic wave is the hydrodynamical mechanism of transporting the CME current; it transforms an initial chiral or electric charge fluctuation into a macroscopic observable asymmetry in the distribution of electric charge \cite{Burnier:2011bf,Hirono:2014oda}. Very recently, the chiral magnetic wave has been derived from the chiral kinetic theory \cite{Stephanov:2014dma}. In particular, in the presence of chirality-flipping transitions, the chiral magnetic wave at frequencies smaller than the transition rate has been found to give rise to a diffusive vector mode  \cite{Stephanov:2014dma}. The chiral magnetic wave thus can provide an appropriate description for the phenomena discussed in section \ref{cme_condsec}.  

\section{CME in heavy ion collisions}\label{cme_hic}

\begin{figure}
\begin{center}
\includegraphics[width=12cm]{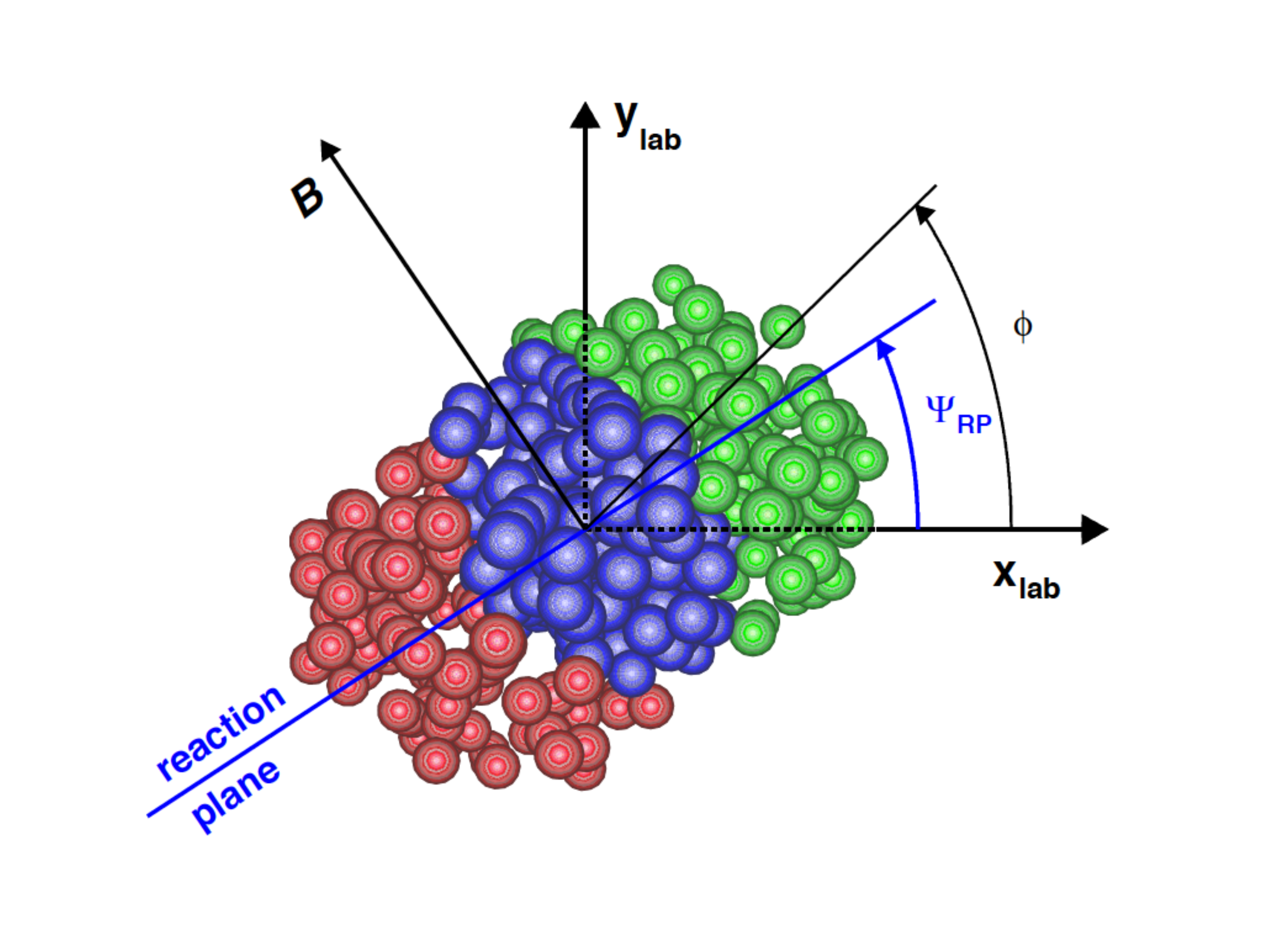}
\end{center}
\caption{Sketch of a heavy ion collision; $\Psi_{RP}$ and $\phi$ are the azimuthal angles of the reaction plane and of a produced hadron; from \cite{Adamczyk:2014mzf}.}
\label{heavy_ion}
\end{figure}

Relativistic heavy ion collisions produce hot QCD matter characterized by strong fluctuations of topological charge. In addition, the colliding ions generate strong magnetic fields $eB \sim {\cal O} (10\ m_\pi^2)$ \cite{Kharzeev:2007jp} that are directed, on the average, orthogonally to the reaction plane, see Fig. \ref{heavy_ion}. 

Let us assume that the produced QCD matter contains a chiral charge produced e.g. due to a local imbalance between the sphaleron and anti-sphaleron transitions discussed in section \ref{qcd_magf}. According to Eq. \ref{cme_current}, there will then be a fluctuation of electric current that will result in a charge asymmetry relative to the reaction plane. This asymmetry will lead to the presence of P-odd harmonics in the angular distribution of the produced positive and negative hadrons:  
\be\label{asym}
\frac{d N_{\pm}}{d\phi} \sim 1 \pm 2 a \sin(\Delta \phi) + ...  .
\ee
with the asymmetry parameter $a \sim 1\%$ \cite{Kharzeev:2004ey}; $\Delta \phi$ is the azimuthal angle relative to the reaction plane, see Fig. \ref{heavy_ion}. 
Since QCD does not violate P and CP globally, the averaged over many events value of the asymmetry is zero, $\langle a \rangle =0$. However the charge asymmetry fluctuates event-by-event, and the signature of CME is the presence of dynamical fluctuations that exceed the statistical ones.  

The number of charged hadron tracks in a single event (although sufficient to determine the reaction plane) is not large enough to allow a statistically sound extraction of the coefficients $a_\pm$, so one has to sum over many events. Because the sign of the charge asymmetry should fluctuate event by event, it was proposed by Voloshin \cite{Voloshin:2004vk} to measure the variable 
\be\label{variable}
\left< \cos(\phi_\alpha + \phi_\beta - 2 \Psi_{RP}) \right> = \left< \cos \Delta \phi_\alpha \cos \Delta \phi_\beta \right> - \left< \sin \Delta \phi_\alpha \sin \Delta \phi_\beta \right> ,
\ee
where the indices $\alpha$ and $\beta$ denote the charge of hadrons, and $\Psi_{RP}$ is the reaction plane angle, see Fig. \ref{heavy_ion}. 
The quantity Eq. \ref{variable} can be measured with a very high precision, and is directly sensitive to the presence of parity--odd fluctuations; it has an added benefit of not being sensitive to the reaction plane--independent backgrounds. There is a  price to pay however -- the observable itself is P-even, and so one has to carefully examine the possible backgrounds; see recent review \cite{Liao:2014ava} for a detailed discussion.

\begin{figure}
\begin{center}
\includegraphics[width=14cm]{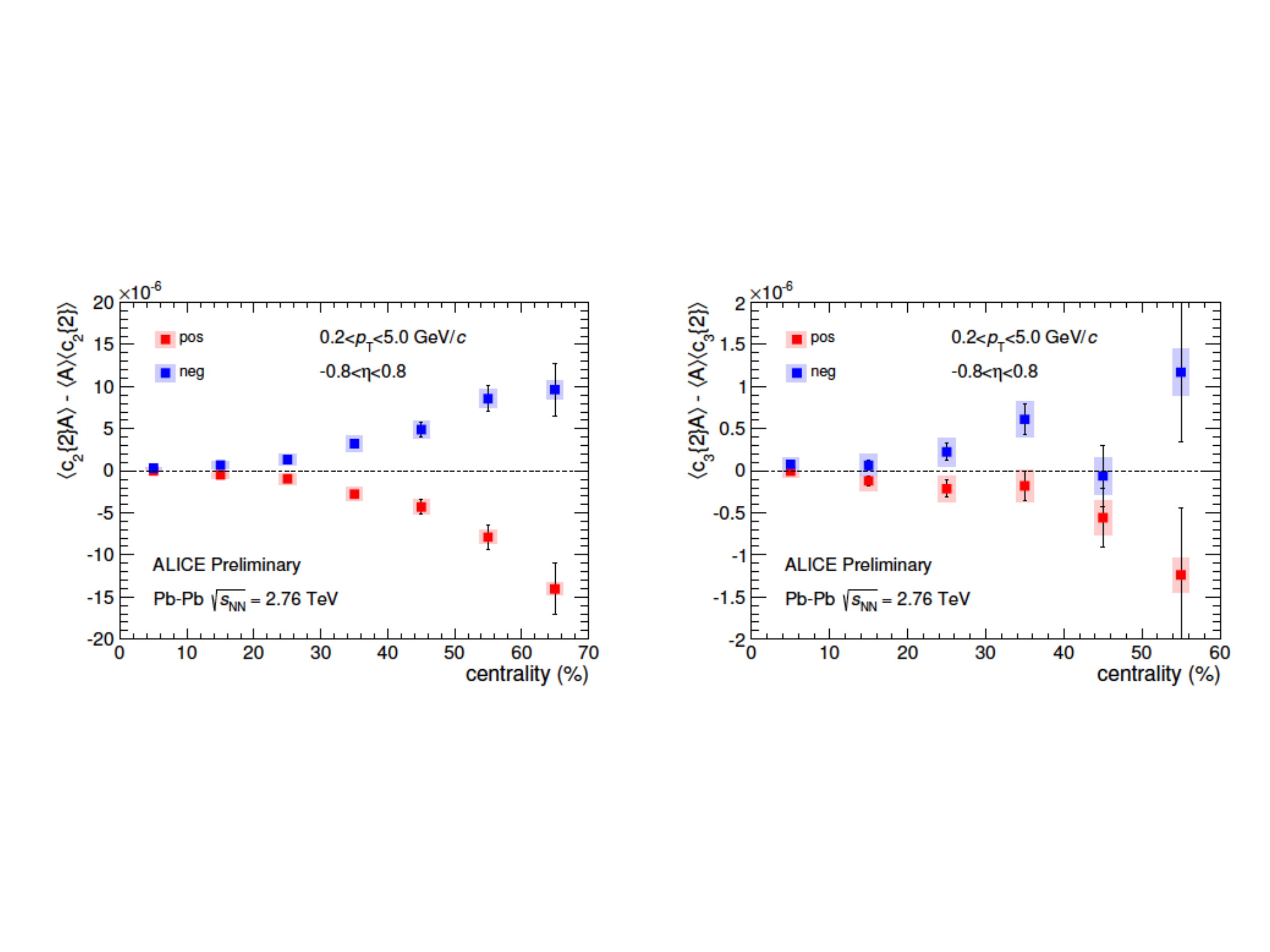}
\end{center}
\vspace{-3cm}
\caption{Charged hadron correlator  as a function of collision centrality measured by the ALICE Collaboration; from \cite{Belmont:2014lta}. Left panel: the second harmonic Eq. \ref{variable}; right panel: the third harmonic of the azimuthal angle distribution.}
\label{alice_cme}
\end{figure}

In 2009, the STAR Collaboration at Relativistic Heavy Ion Collider at BNL reported \cite{Abelev:2009ac,Abelev:2009ad} the measurement of the fluctuation of charge asymmetry Eq. \ref{variable} at $\sqrt{s_{NN}} = 200$ GeV; the order of magnitide of the measured asymmetry was consistent with expectations \cite{Kharzeev:2004ey}. Recently, the ALICE Collaboration Large Hadron Collider at CERN has reported \cite{Belmont:2014lta} the measurement of the charge asymmetry at $\sqrt{s_{NN}} = 2.76$ TeV,  see Fig. \ref{alice_cme}. The magnitude of the observed asymmetry is close to the one previously observed at RHIC -- this is consistent with the fact that magnetic flux through the QCD plasma is approximately energy-independent (magnetic field is proportional to the $\gamma$ factor, and the longitudinal extension of the produced matter at early time is $\sim 1/\gamma$). 

The backgrounds to the measured charge cumulants from local hadron correlations are expected to contribute to both the second harmonic Eq.\ref{variable} and the higher harmonics. Fig. \ref{alice_cme}, right shows the third harmonic as measured by the ALICE Collaboration -- one can see that it is quite small, about an order of magnitude smaller than the second harmonic that is sensitive to CME. 

\begin{figure}
\begin{center}
\includegraphics[width=14cm]{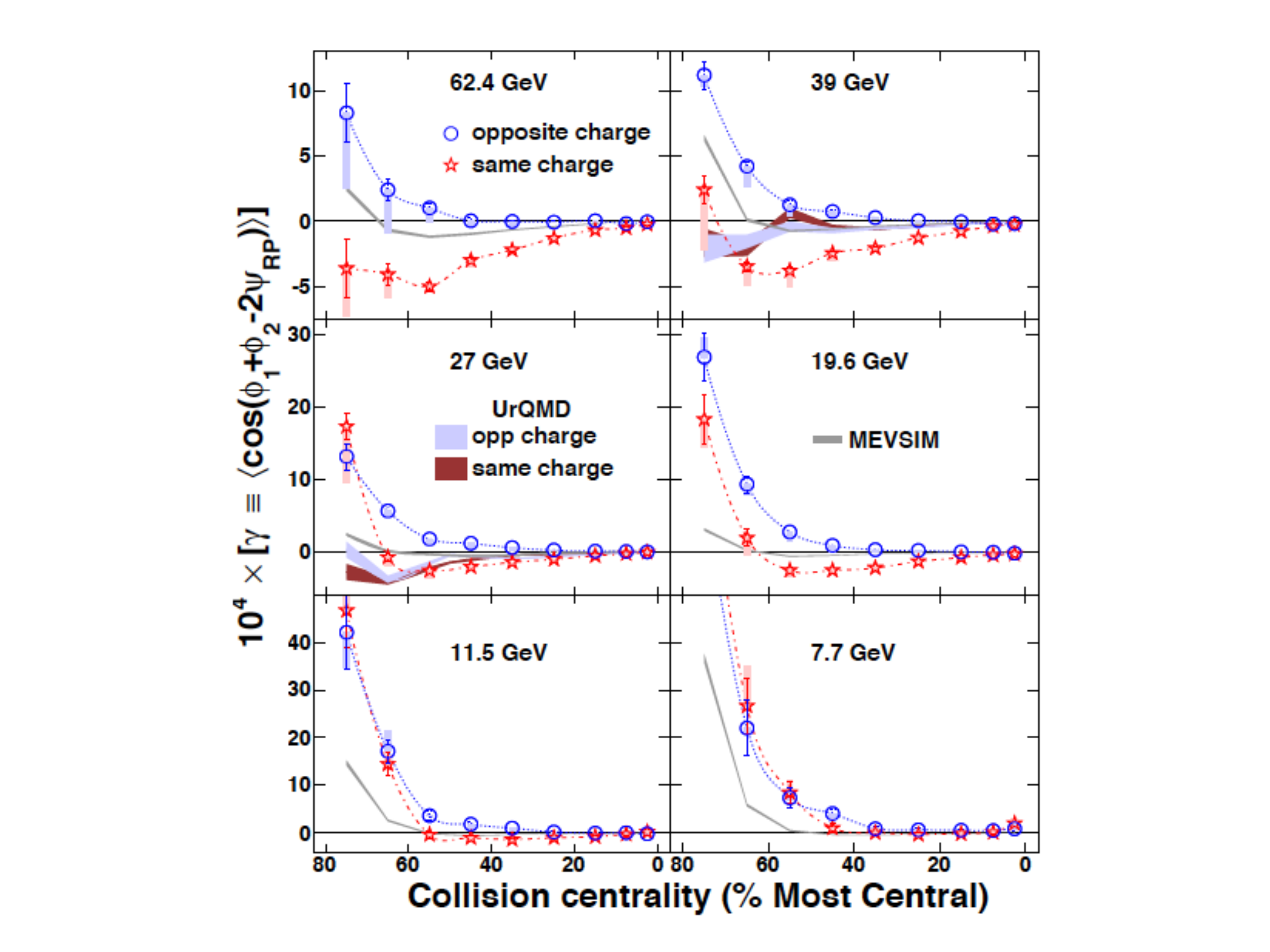}
\end{center}
\caption{The three-point correlator measured by STAR Collaboration in Au-Au collisions at RHIC as a function of centrality at different collision energies; from \cite{Adamczyk:2014mzf}.}
\label{heavy_ion_corr}
\end{figure}

The propagation of the chiral magnetic wave in QCD fluid requires the restoration of chiral symmetry \cite{Kharzeev:2010gd,Burnier:2011bf} -- therefore one expects a reduction in the strength of charge asymmetries when the collision energy decreases below the energy needed to induce the QCD transition.  This is consistent with the results of STAR Collaboration from the beam energy scan at RHIC \cite{Adamczyk:2014mzf}, see Fig. \ref{heavy_ion_corr}.

\begin{figure}
\begin{center}
\includegraphics[width=14cm]{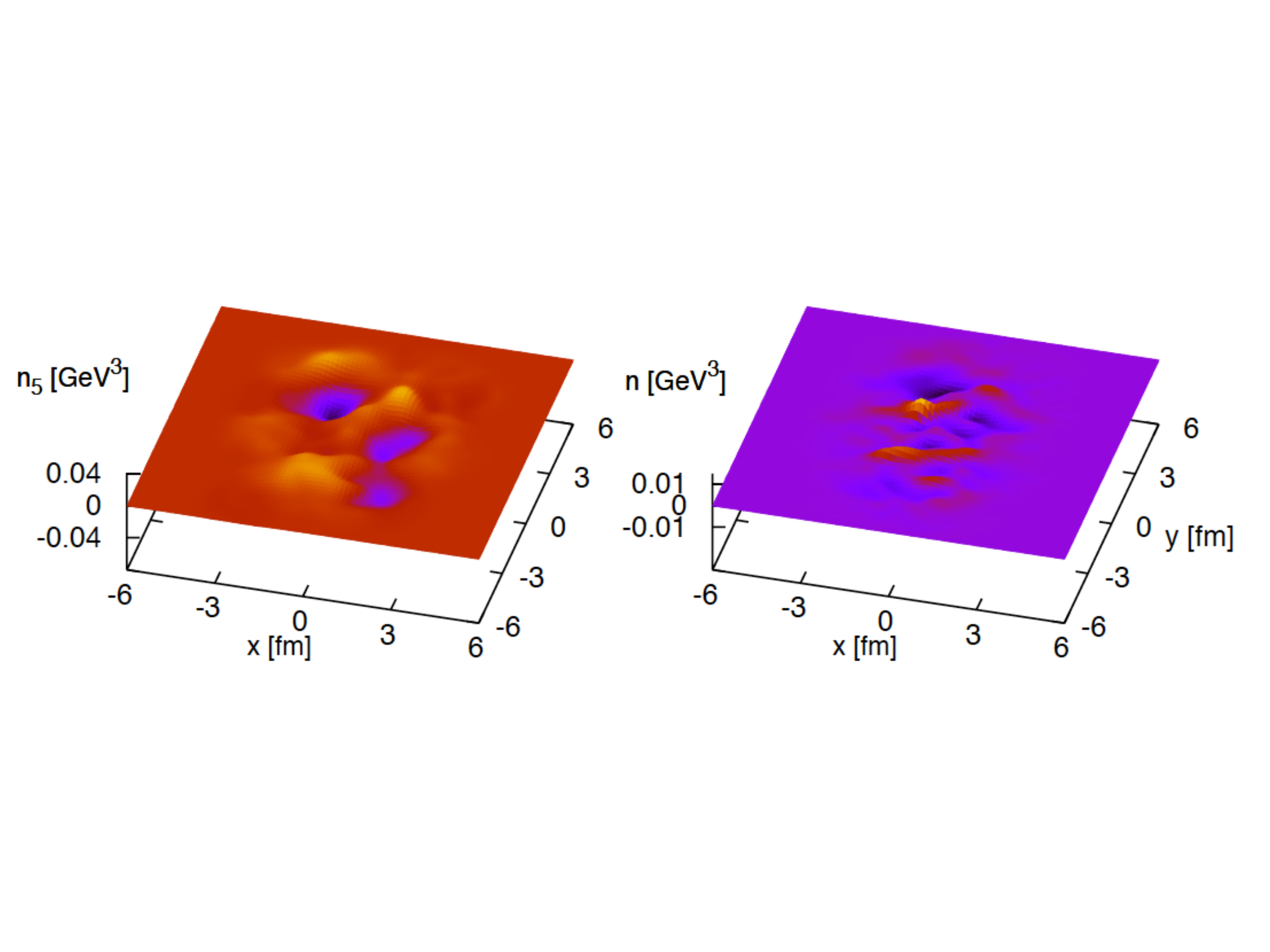}
\end{center}
\vspace{-3cm}
\caption{Distributions of the chiral (left) and electric (right) charge densities in the transverse plane at mid-rapidity and a proper time $\tau = 1.5$fm of a AuAu heavy ion collision at $\sqrt{s} =$ 200 GeV per nucleon as computed  in anomalous hydrodynamics; from \cite{Hirono:2014oda}.}
\label{axial_electric}
\end{figure}

\begin{figure}
\begin{center}
\includegraphics[width=14cm]{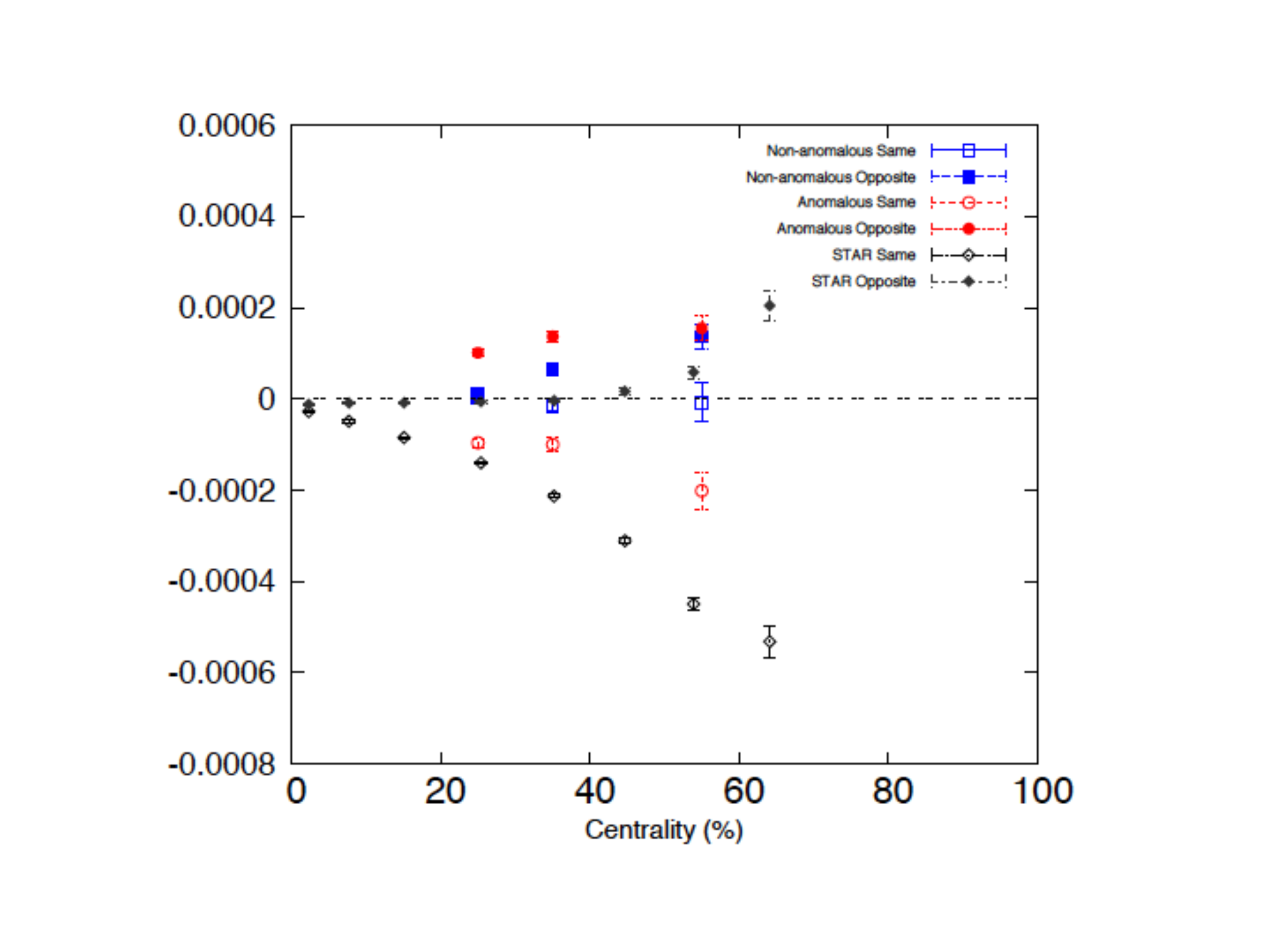}
\end{center}
\vspace{-1cm}
\caption{The three-point correlator $\gamma$ for same and opposite charge hadrons as computed using anomalous hydrodynamics compared to the results from STAR Collaboration in Au-Au collisions at $\sqrt{s} =$ 200 GeV per nucleon; from \cite{Hirono:2014oda}.}
\label{an_hyd_star}
\end{figure}

A fully quantitative theoretical approach to the description of charge asymmetries requires the use of relativistic hydrodynamics that includes the terms arising from the chiral anomaly, supplemented by the initial conditions describing topological fluctuations at the early stage of a heavy collision. The study of that kind has recently been performed in \cite{Hirono:2014oda}; the initial condition is provided by the fluctuating longitudinal ``glasma" fields with 
${{\vec E}}^a {{\vec B}}^a \neq 0$ \cite{Kharzeev:2001ev,Lappi:2006fp,Kharzeev:2005iz}. The snapshot of the resulting chiral and electric charge densities in the QCD fluid is shown in Fig. \ref{axial_electric}. The computed (with and without anomalous terms) charge asymmetries are shown in Fig. \ref{an_hyd_star} in comparison to the STAR experimental results \cite{Abelev:2009ac}. One can see that i) chiral anomaly has a big effect on charge asymmetries and ii) the results of the computation including the anomaly (and thus the CME) agree with the data within a factor of two, possibly leaving room for some background contributions. 

One of the remaining sources of uncertainty in theoretical calculations is the duration of magnetic field in QCD fluid, see \cite{Tuchin:2013ie} for a review. However this uncertainty can be reduced by the study of directed flow of charged hadrons away from mid-rapidity, as  proposed recently in \cite{Gursoy:2014aka}. The magnetic field is also expected to contribute to the photon and dilepton production through the ``magneto-sono-luminescence": the conversion of phonons into real or virtual photons in a magnetic background \cite{Basar:2012bp,Basar:2014swa}. 

\section{CME in condensed matter: the first observation}\label{cme_condsec}

The recent discovery \cite{Borisenko2014,Neupane2014,Liu2014} of Dirac semimetals with chiral quasi-particles opens a fascinating possibility to study the chiral magnetic effect (CME) in condensed matter experiments. Very recently, the first observation of CME was reported through the measurement of magneto-transport in zirconium pentatelluride, ZrTe$_5$ \cite{Li:2014bha}. 

The signature of the CME in Dirac systems in parallel electric and magnetic fields is the positive contribution to the conductivity that has a quadratic dependence on magnetic field 
\cite{Fukushima:2008xe,Son:2012bg,Burkov2014}. This is because the CME current Eq. \ref{cme_current} is proportional to the product of chiral chemical potential and the magnetic field, and the chirality imbalance in Dirac systems is generated dynamically through the chiral anomaly with a rate that is proportional to the product of electric and magnetic fields. As a result, the longitudinal magnetoresistance becomes negative \cite{Son:2012bg,Burkov2014}, with a definite telltale dependence on magnetic field. In the framework of holography, the negative magnetoresistance was recently reproduced in \cite{Landsteiner:2014vua}. Let us explain how this mechanism works in Dirac semimetals in more detail. 

 In the absence of external fields, each Dirac point contains left- and right-handed fermions with equal chemical potentials, $\mu_L = \mu_R = 0$.  If the energy degeneracy between the left- and right-handed fermions gets broken, we can parameterize it by introducing the chiral chemical potential $\mu_5 \equiv (\mu_R - \mu_L)/2$. The corresponding density of chiral charge is then given by \cite{Fukushima:2008xe}
\be\label{chir_density}
\rho_5 = \frac{\mu_5^3}{3 \pi^2 v^3} + \frac{\mu_5}{3 v^3}\ \left(T^2 + \frac{\mu^2}{\pi^2}\right).
\ee

The chiral anomaly in parallel external electric and magnetic fields generates the chiral charge with the rate given by
\be
\frac{d \rho_5}{dt} = \frac{e^2}{4 \pi^2 \hbar^2 c} \vec{E}\cdot \vec{B} .
\ee
The left- and right-handed fermions in Dirac semimetals can mix through chirality-changing scattering, and this process depletes the maximal amount of chiral charge that can be produced. Denoting the chirality-changing scattering time by $\tau_V$, we thus get the equation
\be\label{anomalyeq}
\frac{d \rho_5}{dt} = \frac{e^2}{4 \pi^2 \hbar^2 c} \vec{E}\cdot \vec{B} - \frac{\rho_5}{\tau_V} .
\ee

The solution of 
equation (\ref{anomalyeq}) at $t \gg \tau_V$ is
\be
\rho_5 = \frac{e^2}{4 \pi^2 \hbar^2 c} \vec{E}\cdot \vec{B} \ \tau_V .
\ee
According to (\ref{chir_density}), this leads to a non-zero chiral chemical potential $\mu_5 = (\mu_R - \mu_L)/2$ (assuming that $\mu_5 \ll \mu, T$):
\be\label{chirpot}
\mu_5 = \frac{3}{4} \frac{v^3}{\pi^2} \frac{e^2}{\hbar^2 c}\ \frac{\vec{E}\cdot \vec{B}}{T^2 + \frac{\mu^2}{\pi^2}} \tau_V .
\ee
The formulae (\ref{cme_current}) and (\ref{chirpot}) yield the expression for the CME current: 
\be\label{cme_cur}
J_{\rm CME}^i = \frac{e^2}{\pi \hbar} \ \frac{3}{8} \frac{e^2}{\hbar c} \ \frac{v^3}{\pi^3}\ \frac{\tau_V}{T^2 + \frac{\mu^2}{\pi^2}}\ B^i B^k E^k  \equiv \sigma^{ik}_{\rm CME} \ E^k.
\ee
We see that the CME is described by the conductivity tensor $\sigma^{ik}_{\rm CME} \sim B^i B^k$. When the electric and magnetic fields are parallel, 
the CME conductivity is 
\be\label{cme_cond}
\sigma^{zz}_{\rm CME} = \frac{e^2}{\pi \hbar} \ \frac{3}{8} \frac{e^2}{\hbar c} \ \frac{v^3}{\pi^3}\ \frac{\tau_V}{T^2 + \frac{\mu^2}{\pi^2}}\ B^2.
\ee 
Since the CME current is directed along the electric field, it will affect the measured conductivity -- the total current will be the sum of the Ohmic and CME ones:
\be\label{ohm}
J = J_{\rm Ohm} + J_{\rm CME} = (\sigma_{\rm Ohm} + \sigma_{\rm CME}) \ E,
\ee
where $\sigma_{\rm CME} \equiv \sigma^{zz}_{\rm CME}$. 

If the electric and magnetic fields are parallel, there is no conventional contribution to magnetoresistance induced by the Lorentz force. The magnetoresistance $\rho = 1/\sigma$ following from (\ref{ohm}) is negative, with a characteristic quadratic dependence on magnetic field. 
It is this negative MR with a quadratic dependence on magnetic field that was observed in ZrTe$_5$ \cite{Li:2014bha}. 

ZrTe$_5$ is a layered material that crystallizes in the layered orthorhombic crystal structure. Prismatic ZrTe$_6$ chains run along the crystallographic $a$-axis and are linked along the $c$-axis via zigzag chains of Te atoms forming two-dimensional (2D) layers stacked along the $b$-axis into a crystal. 

\begin{figure}
\begin{center}
\includegraphics[width=14cm]{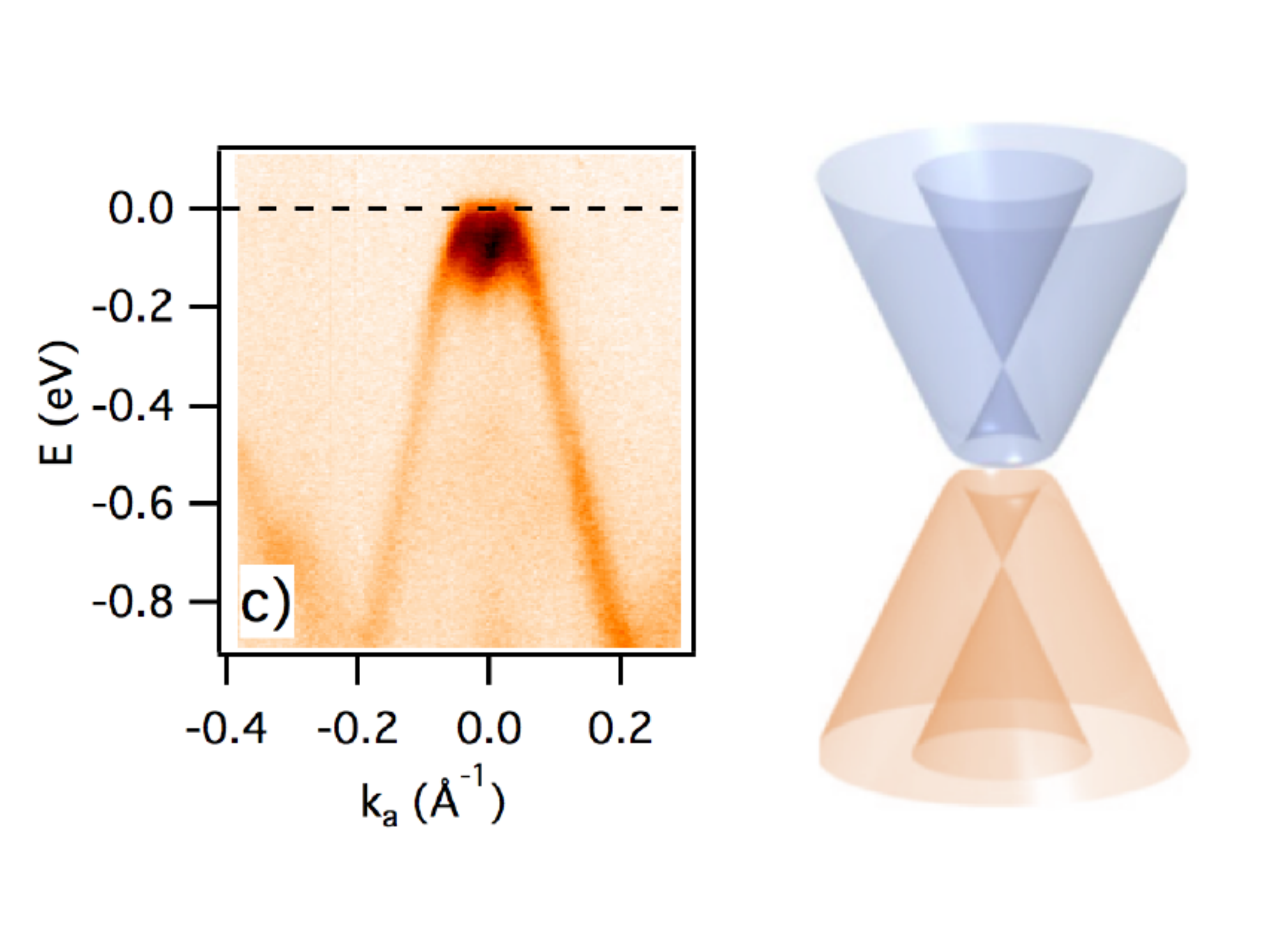}
\end{center}
\vspace{-1cm}
\caption{Electronic structure of ${\rm ZrTe}_5$: the valence band dispersion along the chain direction measured by Angle Resolved Photo Emission Spectroscopy (left); schematic view of the in-plane low energy electronic structure; from \cite{Li:2014bha}.}
\label{zrte_band}
\end{figure}

The requirement for observation of the CME is that a material has a 3D Dirac semimetal-like (zero gap), or semiconductor-like (non-zero gap) electronic structure. Figure \ref{zrte_band}, left shows angle-resolved photoemission spectroscopy (ARPES) data \cite{Li:2014bha} from a freshly cleaved ($a-c$ plane)
 ZrTe$_5$ sample. The states forming the Fermi surface disperse linearly over a large energy range, both along the chain direction (Figure \ref{zrte_band}, left) and perpendicular to it \cite{Li:2014bha}, indicating a Dirac-like dynamics of carriers for the in-plane propagation. The velocity (the slope of dispersion) is very large in both the chain direction, $v_a\simeq 6.4$ eV\AA  ($\simeq c/300$), and perpendicular to it, $v_c\simeq4.5$ eV\AA -- these values are close to the velocity in graphene, $v \simeq c/300$. 
 
 However, the in-plane electronic structure of ZrTe$_5$ cannot be described by a single (anisotropic) cone, especially away from very low energies. The states are doubled - this is because the crystal contains two layers per unit cell - and the simplest description is that the bi-layer splitting creates two cones (bonding-antibonding), separated in energy by $\sim 300$ meV possibly with a small gap at the degeneracies, as shown in Figure \ref{zrte_band}, right, from \cite{Li:2014bha}.

\begin{figure}
\begin{center}
\includegraphics[width=14cm]{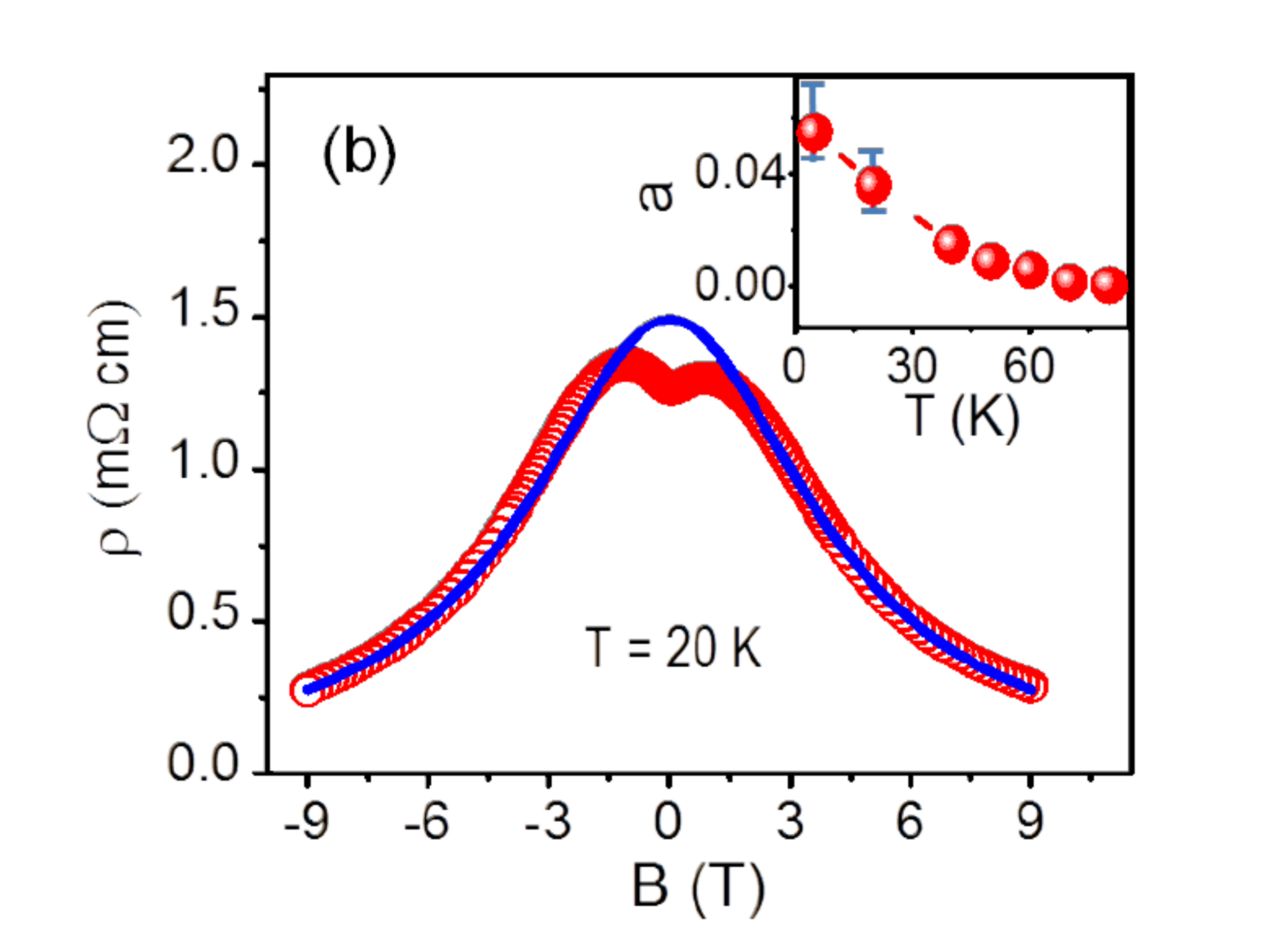}
\end{center}
\vspace{-1cm}
\caption{Longitudinal magnetoresistance of ${\rm ZrTe}_5$:  the experimental measurement (open red points) compared to the prediction from the chiral magnetic effect (blue line); from \cite{Li:2014bha}.}
\label{zrte5}
\end{figure}

A negative magnetoresistance (MR) is observed \cite{Li:2014bha} in ${\rm ZrTe}_5$ for $T\leq100$ K, increasing in magnitude as temperature decreases. The magnetic field dependence of the negative MR can be nicely described by the CME contribution to the electrical conductivity, given by  (\ref{cme_cond}) -- this is illustrated in Fig. \ref{zrte5}, right  for $T=20$ K, from \cite{Li:2014bha}. At 4 Tesla, the CME conductivity is about the same as the zero-field conductivity. At 9T, the CME contribution increases by $\sim 400\%$, resulting in a negative MR that is much stronger than any conventional one reported at an equivalent magnetic field in a non-magnetic material. Inset in Fig. \ref{zrte5} shows the temperature dependence of the pre-factor of $B^2$ which decreases with temperature faster than $1/T$, consistent with the CME.

At very low field, the data show a small cusp-like feature. The origin of this feature is not completely understood, but it probably indicates some form of anti-localization. Similar feature has also been seen in the Dirac semimetal ${\rm Cd_3As_2}$ \cite{Liang}, where an indication of negative magnetoresistance at larger fields has also been reported. It would be interesting to understand the origin of this positive MR at weak fields. 

As pointed out in \cite{Li:2014bha}, the studies of CME can be extended to a broad range of materials since 3D Dirac semimetals often emerge at quantum transitions between normal and topological insulators, including topological crystalline insulators. Dirac semimetals also open new possibilities for photonics since they are expected to possess well pronounced plasmon excitations in the THz frequency range, with universal properties \cite{Kharzeev:2014sba}.

\section{Topology and the Universe}

\subsection{CME in neutron stars}\label{neutron}

Supernova collapse is known to provide a kick to the emerging proto-neutron stars. The velocity 
of the produced neutron stars reaches $v = 1083^{+103}_{-90}$ km/s, as measured by Very Long Baseline Array  \cite{Chatterjee:2005mj} for the pulsar B1508+55. 
  The mechanism of this kick remains a mystery. Several of the proposed explanations involve the interplay of neutrino physics with the magnetic field, e.g. the anisotropy of neutrino emission \cite{Vilenkin:1980fu} or neutrino oscillations biased by magnetic fields \cite{Kusenko:1996sr}.

It has been proposed in \cite{Charbonneau:2009ax} that the neutron star kick can be provided by the chiral magnetic current arising from the asymmetry between left- and right-handed electrons due to the parity violation in weak interactions. The authors estimate that CME throughout the typical age $t \simeq 40,000$ years of a neutron star can provide it a velocity on the order of $v \sim 1000$ km/s.  

A different CME-related mechanism of the neutron star kick was proposed in \cite{Gorbar:2009bm}.
The idea of  \cite{Gorbar:2009bm} is based on the ``chiral separation effect" that is dual to the CME -- it is a flow of the axial current induced by the baryon chemical potential. Because neutrino scattering can tell the difference between left- and right-handed fermions, this flow of axial current induces an anisotropy in neutrino emission and thus the desired kick of the neutron star. 

It would be interesting to explore the propagation of the axial current using the anomalous hydrodynamics, in analogy to previous studies of the ``chiral magnetic wave" \cite{Kharzeev:2010gd,Burnier:2011bf} in heavy ion collisions. 
The first steps in using anomalous hydrodynamics have already been made \cite{Kaminski:2014jda}. In addition to magnetic field, the rotation of neutron stars also has to be taken into account since they bring in the effects related to gravitational anomaly \cite{Shaverin:2014xya}. 

Another long-standing problem in the physics of neutron stars is the origin of their strong magnetic fields. The Maxwell-Chern-Simons theory that provides \cite{Kharzeev:2009fn} an effective low energy theory of CME  in the presence of chiral chemical potential $\mu_5$ displays a ``chiral magnetic instability" \cite{Redlich:1984md,Tsokos:1985ny,Joyce:1997uy,Frohlich:2000en} of the magnetic field at small momenta $k \leq e^2 \mu_5/(2 \pi)$. This instability  leads to the possibility of producing helical magnetic fields characterized by non-zero value of magnetic helicity (Chern-Simons 3-form) $\int d^3x\ \vec{A} \cdot \vec{B}$. During the collapse of neutron stars, left-handed electrons are captured by protons due to weak interactions. This induces a chiral imbalance that through the chiral magnetic instability can power the growth of a helical magnetic field, as proposed in \cite{Ohnishi:2014uea}. 

A more detailed investigation \cite{Grabowska:2014efa} however revealed the important role of electron mass that dampens the chiral magnetic instability, at least in the formulation of \cite{Ohnishi:2014uea}. It remains to be seen if the chiral magnetic instability can be saved in a more elaborate treatment taking into account the energy balance in magnetic and gravitational fields of the star\footnote{I thank S. Reddy for a stimulating discussion of this issue.}.

\subsection{CME and the primordial magnetic fields}\label{cme_prim}

An intriguing application of CME that stimulated some of the early work \cite{Vilenkin:1982pn,Joyce:1997uy,Giovannini:1997gp,Frohlich:2000en} as described in review \cite{Kharzeev:2013ffa} is the generation of primordial magnetic fields in the Early Universe. The chirality imbalance in the primordial electroweak plasma can be converted by the chiral anomaly into a helical magnetic field configuration with non-zero Chern-Simons number. 

To explain the presence of relatively strong magnetic fields and magnetic helicity in the present day Universe, one has to understand  the transfer of magnetic helicity from small to large scales \cite{Boyarsky:2011uy}, study of the conversion of the electroweak plasma into a horizon-scale helical magnetic field \cite{Boyarsky:2012ex}, and the realization that leptogenesis can give rise to right-handed helical magnetic field that is coherent on astrophysical length scales \cite{Long:2013tha}. Similar ideas have been developed in the series of papers \cite{Semikoz:2007ti,Semikoz:2009ye,Semikoz:2012ka}. The chiral vortical effect has been found to lead to the production of the helical magnetic field in the turbulent electroweak plasma \cite{Tashiro:2012mf}, whereas the CME amplifies the growth of the field. 

A particularly intriguing option for the Universe is the inflationary expansion driven by a pseudo-scalar inflaton -- for example, a pseudo-Nambu-Goldstone boson \cite{Freese:1990rb} or an axion \cite{Kim:2004rp}. In this case the entire Universe would be in a parity-odd state, and the coupling to gauge fields would induce CME currents on cosmological scales. In particular, this cosmological parity violation would be imprinted in the Cosmic Microwave Background, as discussed recently in \cite{Sorbo:2011rz,Barnaby:2011vw,Alexander:2011hz,Zhitnitsky:2013pna,Zhitnitsky:2014aja,Ferreira:2014zia,Bartolo:2014hwa}. 

\section{Outlook}

Topological field configurations define many fundamental 
 properties of the physical world, including spontaneous breaking of chiral symmetry and confinement in QCD. As argued throughout this review, coherent magnetic fields provide an ideal probe of QCD topology. 
For the investigation of non-perturbative dynamics of QCD, external magnetic fields may well play a role similar to the one played by deep-inelastic scattering in establishing the short distance perturbative behavior of quarks and gluons.  

Strong magnetic fields are accessible in heavy ion collisions and in numerical simulations on the lattice; this opens unique possibilities for testing theoretical ideas about QCD topology and gaining a deeper understanding of strong interactions. The presence of strong magnetic fields (up to $\sim 10^{15}\ \mathrm{G}$) in neutron stars provides an access to magnetized cold dense QCD matter. The knowledge gained about topological effects in non-Abelian plasmas can advance the understanding of the baryon asymmetry and magnetic helicity in the Universe. 

Finally, the recent discovery of Dirac semimetals and the observation of the chiral magnetic effect in ZrTe$_5$ open an exciting possibility to study 3D topological field-theoretic phenomena in condensed matter systems, with a potential for a wide array of applications.

\vskip0.3cm
I thank M. Chernodub, M. D'Elia, G. Endrodi, Z. Xu and the STAR Collaboration for the permission to use graphics from their papers and helpful comments on the manuscript. This work was supported in part by the U.S. Department of Energy under Contracts
DE-FG-88ER40388 and DE-SC0012704.


\begin{thebibliography}{99}

\bibitem{Belavin:1975fg} 
  A.~A.~Belavin, A.~M.~Polyakov, A.~S.~Schwartz and Y.~S.~Tyupkin,
  Phys.\ Lett.\ B {\bf 59}, 85 (1975).
  
\bibitem{Callan:1976je} 
  C.~G.~Callan, Jr., R.~F.~Dashen and D.~J.~Gross,
  Phys.\ Lett.\ B {\bf 63}, 334 (1976).
 
\bibitem{'tHooft:1976fv} 
  G.~'t Hooft,
  Phys.\ Rev.\ D {\bf 14}, 3432 (1976)
  [Erratum-ibid.\ D {\bf 18}, 2199 (1978)].
  
\bibitem{Schafer:1996wv} 
  T.~Schafer and E.~V.~Shuryak,
  Rev.\ Mod.\ Phys.\  {\bf 70}, 323 (1998)
  [hep-ph/9610451].



\bibitem{Kharzeev:2013jha} 
  D.~Kharzeev, K.~Landsteiner, A.~Schmitt and H.~U.~Yee (Eds),
  Lect.\ Notes Phys.\  {\bf 871}, 1-624 (2013).

\bibitem{Kharzeev:2013ffa} 
  D.~E.~Kharzeev,
  Prog.\ Part.\ Nucl.\ Phys.\  {\bf 75}, 133 (2014)
  [arXiv:1312.3348 [hep-ph]].
 
\bibitem{Shovkovy:2012zn} 
  I.~A.~Shovkovy,
  Lect.\ Notes Phys.\  {\bf 871}, 13 (2013)
  [arXiv:1207.5081 [hep-ph]].

\bibitem{Andersen:2014xxa} 
  J.~O.~Andersen, W.~R.~Naylor and A.~Tranberg,
  arXiv:1411.7176 [hep-ph].

\bibitem{Rebhan:2014rxa} 
  A.~Rebhan,
  arXiv:1410.8858 [hep-th].
 
\bibitem{Shuryak:2014zxa} 
  E.~Shuryak,
  arXiv:1412.8393 [hep-ph].

\bibitem{Kharzeev:2007jp} 
  D.~E.~Kharzeev, L.~D.~McLerran and H.~J.~Warringa,
  Nucl.\ Phys.\ A {\bf 803}, 227 (2008)
  [arXiv:0711.0950 [hep-ph]].

\bibitem{Tuchin:2013ie} 
  K.~Tuchin,
  Adv.\ High Energy Phys.\  {\bf 2013}, 490495 (2013)
  [arXiv:1301.0099].
 
 \bibitem{Ong}
 Q. D. Gibson, L. M. Schoop, L. Muechler, L. S. Xie, M. Hirschberger, N. P. Ong, R. Car, R. J. Cava,	
 arXiv:1411.0005
 
\bibitem{Li:2014bha} 
  Q.~Li, D.~E.~Kharzeev, C.~Zhang, Y.~Huang, I.~Pletikosic, A.~V.~Fedorov, R.~D.~Zhong, J.~A.~Schneeloch, G.~D.~Gu and T.~Valla,
  arXiv:1412.6543 [cond-mat.str-el].

\bibitem{Witten:1998uka} 
  E.~Witten,
  Phys.\ Rev.\ Lett.\  {\bf 81}, 2862 (1998)
  [hep-th/9807109].


\bibitem{Bonati:2013tt} 
  C.~Bonati, M.~DÕElia, H.~Panagopoulos and E.~Vicari,
  Phys.\ Rev.\ Lett.\  {\bf 110}, no. 25, 252003 (2013)
  [arXiv:1301.7640 [hep-lat]].

\bibitem{Bonati:2013dza} 
  C.~Bonati, M.~D'Elia, H.~Panagopoulos and E.~Vicari,
  PoS LATTICE {\bf 2013}, 136 (2014)
  [arXiv:1309.6059 [hep-lat]].

\bibitem{Kharzeev:1998kz} 
  D.~Kharzeev, R.~D.~Pisarski and M.~H.~G.~Tytgat,
  Phys.\ Rev.\ Lett.\  {\bf 81}, 512 (1998)
  [hep-ph/9804221].

\bibitem{Gross:1980br} 
  D.~J.~Gross, R.~D.~Pisarski and L.~G.~Yaffe,
  Rev.\ Mod.\ Phys.\  {\bf 53}, 43 (1981).

\bibitem{Shuryak:1978yk} 
  E.~V.~Shuryak,
  Phys.\ Lett.\ B {\bf 79}, 135 (1978).

\bibitem{Pisarski:1980md} 
  R.~D.~Pisarski and L.~G.~Yaffe,
  Phys.\ Lett.\ B {\bf 97}, 110 (1980).

 
  
  \bibitem{Klinkhamer:1984di}
N.~S.~Manton,
  Phys.\ Rev.\  D {\bf 28}, 2019 (1983);
  F.~R.~Klinkhamer and N.~S.~Manton,
  Phys.\ Rev.\  D {\bf 30}, 2212 (1984).
  
\bibitem{Kuzmin:1985mm}
  V.~A.~Kuzmin, V.~A.~Rubakov and M.~E.~Shaposhnikov,
  Phys.\ Lett.\  B {\bf 155}, 36 (1985).
 
\bibitem{Rubakov:1996vz}
  V.~A.~Rubakov and M.~E.~Shaposhnikov,
  Usp.\ Fiz.\ Nauk {\bf 166}, 493 (1996)
  [Phys.\ Usp.\  {\bf 39}, 461 (1996)]
  [arXiv:hep-ph/9603208].

\bibitem{McLerran:1990de}
  L.~D.~McLerran, E.~Mottola and M.~E.~Shaposhnikov,
  Phys.\ Rev.\  D {\bf 43}, 2027 (1991).
 
\bibitem{Arnold:1996dy}
  P.~Arnold, D.~Son and L.~G.~Yaffe,
  Phys.\ Rev.\  D {\bf 55}, 6264 (1997)
  [arXiv:hep-ph/9609481].
  
\bibitem{Huet:1996sh}
  P.~Huet and D.~T.~Son,
  Phys.\ Lett.\  B {\bf 393}, 94 (1997)
  [arXiv:hep-ph/9610259].
 
\bibitem{Bodeker:1998hm}
  D.~Bodeker,
  Phys.\ Lett.\  B {\bf 426}, 351 (1998)
  [arXiv:hep-ph/9801430].
  
\bibitem{Moore:1997sn}
  G.~D.~Moore, C.~r.~Hu and B.~Muller,
  Phys.\ Rev.\  D {\bf 58}, 045001 (1998)
  [arXiv:hep-ph/9710436];
 D.~Bodeker, G.~D.~Moore and K.~Rummukainen,
  Phys.\ Rev.\  D {\bf 61}, 056003 (2000).

\bibitem{Moore:2010jd} 
  G.~D.~Moore and M.~Tassler,
  JHEP {\bf 1102}, 105 (2011)
  [arXiv:1011.1167 [hep-ph]].

\bibitem{AdS-CFT}
 J.~M.~Maldacena,
  Adv.\ Theor.\ Math.\ Phys.\  {\bf 2} (1998) 231
  [Int.\ J.\ Theor.\ Phys.\  {\bf 38} (1999) 1113]
  [arXiv:hep-th/9711200];\,\,\,
  
 \bibitem{AdS-CFT1}
 
S.~S.~Gubser, I.~R.~Klebanov and A.~M.~Polyakov,
  Phys.\ Lett.\  B {\bf 428} (1998) 105
  [arXiv:hep-th/9802109];\,\,\,
  
 \bibitem{AdS-CFT2}

E.~Witten,
  Adv.\ Theor.\ Math.\ Phys.\  {\bf 2} (1998) 505
  [arXiv:hep-th/9803131].

\bibitem{Son:2002sd}
  D.~T.~Son and A.~O.~Starinets,
  JHEP {\bf 0209}, 042 (2002)
  [arXiv:hep-th/0205051].

\bibitem{Bazavov:2014pvz} 
  A.~Bazavov {\it et al.}  [HotQCD Collaboration],
  Phys.\ Rev.\ D {\bf 90}, no. 9, 094503 (2014)
  [arXiv:1407.6387 [hep-lat]].

\bibitem{Gursoy:2012bt} 
  U.~Gursoy, I.~Iatrakis, E.~Kiritsis, F.~Nitti and A.~O'Bannon,
  JHEP {\bf 1302}, 119 (2013)
  [arXiv:1212.3894 [hep-th]].

\bibitem{Gursoy:2010fj} 
  U.~Gursoy, E.~Kiritsis, L.~Mazzanti, G.~Michalogiorgakis and F.~Nitti,
  Lect.\ Notes Phys.\  {\bf 828}, 79 (2011)
  [arXiv:1006.5461 [hep-th]].

\bibitem{Gibbons:1995vg}
  G.~W.~Gibbons, M.~B.~Green and M.~J.~Perry,
  Phys.\ Lett.\  B {\bf 370}, 37 (1996)
  [arXiv:hep-th/9511080].

\bibitem{Green:1997tv}
  M.~B.~Green and M.~Gutperle,
  Nucl.\ Phys.\  B {\bf 498}, 195 (1997)
  [arXiv:hep-th/9701093].



\bibitem{Kharzeev:2009pa} 
  D.~E.~Kharzeev and E.~M.~Levin,
  JHEP {\bf 1001}, 046 (2010)
  [arXiv:0910.3355 [hep-ph]].

\bibitem{Kharzeev:1999vh} 
  D.~Kharzeev and E.~Levin,
  Nucl.\ Phys.\ B {\bf 578}, 351 (2000)
  [hep-ph/9912216].
  
\bibitem{Kharzeev:2000ef} 
  D.~E.~Kharzeev, Y.~V.~Kovchegov and E.~Levin,
  Nucl.\ Phys.\ A {\bf 690}, 621 (2001)
  [hep-ph/0007182].
  
\bibitem{Shuryak:2000df} 
  E.~V.~Shuryak and I.~Zahed,
  Phys.\ Rev.\ D {\bf 62}, 085014 (2000)
  [hep-ph/0005152].

\bibitem{Nowak:2000de} 
  M.~A.~Nowak, E.~V.~Shuryak and I.~Zahed,
  Phys.\ Rev.\ D {\bf 64}, 034008 (2001)
  [hep-ph/0012232].

\bibitem{Fukushima:2008xe} 
  K.~Fukushima, D.~E.~Kharzeev and H.~J.~Warringa,
  Phys.\ Rev.\ D {\bf 78}, 074033 (2008)
  [arXiv:0808.3382 [hep-ph]].

\bibitem{Fukushima:2010fe} 
  K.~Fukushima, M.~Ruggieri and R.~Gatto,
  Phys.\ Rev.\ D {\bf 81}, 114031 (2010)
  [arXiv:1003.0047 [hep-ph]].

\bibitem{Chernodub:2011fr} 
  M.~N.~Chernodub and A.~S.~Nedelin,
  Phys.\ Rev.\ D {\bf 83}, 105008 (2011)
  [arXiv:1102.0188 [hep-ph]].

\bibitem{Gatto:2011wc} 
  R.~Gatto and M.~Ruggieri,
  Phys.\ Rev.\ D {\bf 85}, 054013 (2012)
  [arXiv:1110.4904 [hep-ph]].
  
  
\bibitem{Stephanov:1998dy} 
  M.~A.~Stephanov, K.~Rajagopal and E.~V.~Shuryak,
  Phys.\ Rev.\ Lett.\  {\bf 81}, 4816 (1998)
  [hep-ph/9806219].
  
\bibitem{Braguta:2014ira} 
  V.~V.~Braguta, V.~A.~Goy, E.-M.~Ilgenfritz, A.~Y.~Kotov, A.~V.~Molochkov, M.~Muller-Preussker, B.~Petersson and A.~Schreiber,
  arXiv:1411.5174 [hep-lat].



\bibitem{Gusynin:1994re} 
  V.~P.~Gusynin, V.~A.~Miransky and I.~A.~Shovkovy,
  Phys.\ Rev.\ Lett.\  {\bf 73}, 3499 (1994)
  [Erratum-ibid.\  {\bf 76}, 1005 (1996)]
  [hep-ph/9405262].
 
 
\bibitem{Aleiner:2007va} 
  I.~L.~Aleiner, D.~E.~Kharzeev and A.~M.~Tsvelik,
  Phys.\ Rev.\ B {\bf 76}, 195415 (2007)
  [arXiv:0708.0394 [cond-mat.mes-hall]].
  
\bibitem{Shushpanov:1997sf} 
  I.~A.~Shushpanov and A.~V.~Smilga,
  Phys.\ Lett.\ B {\bf 402}, 351 (1997)
  [hep-ph/9703201].

\bibitem{Klevansky:1989vi} 
  S.~P.~Klevansky and R.~H.~Lemmer,
  Phys.\ Rev.\ D {\bf 39}, 3478 (1989).

\bibitem{Buividovich:2008wf} 
  P.~V.~Buividovich, M.~N.~Chernodub, E.~V.~Luschevskaya and M.~I.~Polikarpov,
  Phys.\ Lett.\ B {\bf 682}, 484 (2010)
  [arXiv:0812.1740 [hep-lat]].

\bibitem{Bali:2011qj} 
  G.~S.~Bali, F.~Bruckmann, G.~Endrodi, Z.~Fodor, S.~D.~Katz, S.~Krieg, A.~Schafer and K.~K.~Szabo,
  JHEP {\bf 1202}, 044 (2012)
  [arXiv:1111.4956 [hep-lat]].


\bibitem{Bruckmann:2013oba}
  F.~Bruckmann, G.~Endrodi and T.~G.~Kovacs,
  JHEP {\bf 1304} (2013) 112
  [arXiv:1303.3972 [hep-lat]].

\bibitem{Bali:2014kia} 
  G.~S.~Bali, F.~Bruckmann, G.~Endršdi, S.~D.~Katz and A.~SchŠfer,
  JHEP {\bf 1408}, 177 (2014)
  [arXiv:1406.0269 [hep-lat]].

\bibitem{Fraga:2012rr} 
  E.~S.~Fraga,
  Lect.\ Notes Phys.\  {\bf 871}, 121 (2013)
  [arXiv:1208.0917 [hep-ph]].


  
  


  
  

\bibitem{Basar:2011by} 
  G.~Basar, G.~V.~Dunne and D.~E.~Kharzeev,
  Phys.\ Rev.\ D {\bf 85}, 045026 (2012)
  [arXiv:1112.0532 [hep-th]].
  
\bibitem{Buividovich:2009wi} 
  P.~V.~Buividovich, M.~N.~Chernodub, E.~V.~Luschevskaya and M.~I.~Polikarpov,
  Phys.\ Rev.\ D {\bf 80}, 054503 (2009)
  [arXiv:0907.0494 [hep-lat]].
  
\bibitem{Buividovich:2009my} 
  P.~V.~Buividovich, M.~N.~Chernodub, E.~V.~Luschevskaya and M.~I.~Polikarpov,
  Phys.\ Rev.\ D {\bf 81}, 036007 (2010)
  [arXiv:0909.2350 [hep-ph]].
  
\bibitem{Buividovich:2010tn} 
  P.~V.~Buividovich, M.~N.~Chernodub, D.~E.~Kharzeev, T.~Kalaydzhyan, E.~V.~Luschevskaya and M.~I.~Polikarpov,
  Phys.\ Rev.\ Lett.\  {\bf 105}, 132001 (2010)
  [arXiv:1003.2180 [hep-lat]].
  
\bibitem{Abramczyk:2009gb} 
  M.~Abramczyk, T.~Blum, G.~Petropoulos and R.~Zhou,
  PoS LAT {\bf 2009}, 181 (2009)
  [arXiv:0911.1348 [hep-lat]].
  



  
\bibitem{Bali:2014vja} 
  G.~S.~Bali, F.~Bruckmann, G.~Endršdi, Z.~Fodor, S.~D.~Katz and A.~SchŠfer,
  JHEP {\bf 1404}, 129 (2014)
  [arXiv:1401.4141 [hep-lat]].


\bibitem{Basar:2012gh} 
  G.~Basar and D.~E.~Kharzeev,
  Phys.\ Rev.\ D {\bf 85}, 086012 (2012)
  [arXiv:1202.2161 [hep-th]].
  
\bibitem{D'Hoker:2009mm}
  E.~D'Hoker, P.~Kraus,
  JHEP {\bf 0910}, 088 (2009).

\bibitem{D'Hoker:2012ih} 
  E.~D'Hoker and P.~Kraus,
  Lect.\ Notes Phys.\  {\bf 871}, 469 (2013)
  [arXiv:1208.1925 [hep-th]].



  
\bibitem{Kharzeev:2004ey} 
  D.~Kharzeev,
  Phys.\ Lett.\ B {\bf 633}, 260 (2006)
  [hep-ph/0406125].


\bibitem{Kharzeev:2007tn} 
  D.~Kharzeev and A.~Zhitnitsky,
  Nucl.\ Phys.\ A {\bf 797}, 67 (2007)
  [arXiv:0706.1026 [hep-ph]].

\bibitem{Kharzeev:2009fn} 
  D.~E.~Kharzeev,
  Annals Phys.\  {\bf 325}, 205 (2010)
  [arXiv:0911.3715 [hep-ph]].


\bibitem{Son:2009tf} 
  D.~T.~Son and P.~Surowka,
  Phys.\ Rev.\ Lett.\  {\bf 103}, 191601 (2009)
  [arXiv:0906.5044 [hep-th]].

\bibitem{Zakharov:2012vv} 
  V.~I.~Zakharov,
  Lect.\ Notes Phys.\  {\bf 871}, 295 (2013)
  [arXiv:1210.2186 [hep-ph]].
  
\bibitem{Kharzeev:2011ds} 
  D.~E.~Kharzeev and H.~-U.~Yee,
  Phys.\ Rev.\ D {\bf 84}, 045025 (2011)
  [arXiv:1105.6360 [hep-th]].

\bibitem{Kharzeev:2010gd} 
  D.~E.~Kharzeev and H.~-U.~Yee,
  Phys.\ Rev.\ D {\bf 83}, 085007 (2011)
  [arXiv:1012.6026 [hep-th]].
  
    
    
\bibitem{Newman:2005hd} 
  G.~M.~Newman,
  JHEP {\bf 0601}, 158 (2006)
  [hep-ph/0511236].

  
\bibitem{Burnier:2011bf} 
  Y.~Burnier, D.~E.~Kharzeev, J.~Liao and H.~U.~Yee,
  Phys.\ Rev.\ Lett.\  {\bf 107}, 052303 (2011)
  [arXiv:1103.1307 [hep-ph]].
 
\bibitem{Hirono:2014oda} 
  Y.~Hirono, T.~Hirano and D.~E.~Kharzeev,
  arXiv:1412.0311 [hep-ph].
  
\bibitem{Stephanov:2014dma} 
  M.~Stephanov, H.~U.~Yee and Y.~Yin,
  arXiv:1501.00222 [hep-th].
  
\bibitem{Voloshin:2004vk} 
  S.~A.~Voloshin,
  Phys.\ Rev.\ C {\bf 70}, 057901 (2004)
  [hep-ph/0406311].

\bibitem{Liao:2014ava} 
  J.~Liao,
  arXiv:1401.2500 [hep-ph].

\bibitem{Abelev:2009ac} 
  B.~I.~Abelev {\it et al.}  [STAR Collaboration],
  Phys.\ Rev.\ Lett.\  {\bf 103}, 251601 (2009)
  [arXiv:0909.1739 [nucl-ex]].

\bibitem{Abelev:2009ad} 
  B.~I.~Abelev {\it et al.}  [STAR Collaboration],
  Phys.\ Rev.\ C {\bf 81}, 054908 (2010)
  [arXiv:0909.1717 [nucl-ex]].


\bibitem{Belmont:2014lta} 
  R.~Belmont [ALICE Collaboration],
  arXiv:1408.1043 [nucl-ex].

\bibitem{Adamczyk:2014mzf} 
  L.~Adamczyk {\it et al.}  [STAR Collaboration],
  Phys.\ Rev.\ Lett.\  {\bf 113}, 052302 (2014)
  [arXiv:1404.1433 [nucl-ex]].


\bibitem{Kharzeev:2001ev} 
  D.~Kharzeev, A.~Krasnitz and R.~Venugopalan,
  Phys.\ Lett.\ B {\bf 545}, 298 (2002)
  [hep-ph/0109253].
  
\bibitem{Lappi:2006fp} 
  T.~Lappi and L.~McLerran,
  Nucl.\ Phys.\ A {\bf 772}, 200 (2006)
  [hep-ph/0602189].
 
 
\bibitem{Kharzeev:2005iz} 
  D.~Kharzeev and K.~Tuchin,
  Nucl.\ Phys.\ A {\bf 753}, 316 (2005)
  [hep-ph/0501234].

 
  

  
     
  
  
        
   
 
 
 

\bibitem{Gursoy:2014aka} 
  U.~Gursoy, D.~Kharzeev and K.~Rajagopal,
  Phys.\ Rev.\ C {\bf 89}, 054905 (2014)
  [arXiv:1401.3805 [hep-ph]].

\bibitem{Basar:2012bp} 
  G.~Basar, D.~Kharzeev and V.~Skokov,
  Phys.\ Rev.\ Lett.\  {\bf 109}, 202303 (2012)
  [arXiv:1206.1334 [hep-ph]].
  
\bibitem{Basar:2014swa} 
  G.~Basar, D.~E.~Kharzeev and E.~V.~Shuryak,
  Phys.\ Rev.\ C {\bf 90}, 014905 (2014)
  [arXiv:1402.2286 [hep-ph]].

\bibitem{Borisenko2014}
S.~Borisenko, Q.~Gibson, D.~Evtushinsky, V.~Zabolotnyy, B.~B\"uchner, and R.~J.
  Cava, 
Phys. Rev. Lett., \textbf{113}, 027603 (2014).

\bibitem{Neupane2014}
M.~Neupane, S.-Y. Xu, R.~Sankar, N.~Alidoust, G.~Bian, C.~Liu, I.~Belopolski,
  T.-R. Chang, H.-T. Jeng, H.~Lin, et~al, 
Nature communications, \textbf{5}, 3786 (2014).

\bibitem{Liu2014}
Z.~K. Liu, J.~Jiang, B.~Zhou, Z.~J. Wang, Y.~Zhang, H.~M. Weng, D.~Prabhakaran,
  S.-K. Mo, H.~Peng, P.~Dudin, T.~Kim, M.~Hoesch, Z.~Fang, X.~Dai, Z.~X. Shen,
  D.~L. Feng, Z.~Hussain, and Y.~L. Chen, 
Nature materials, \textbf{13}, 677 (2014).
  
\bibitem{Son:2012bg} 
  D.~T.~Son and B.~Z.~Spivak,
  Phys.\ Rev.\ B {\bf 88}, 104412 (2013)
  [arXiv:1206.1627 [cond-mat.mes-hall]].
  
  \bibitem{Burkov2014}
A.~A. Burkov, 
Phys.\ Rev.\ Lett., \textbf{113}, 247203 (2014).

\bibitem{Landsteiner:2014vua} 
  K.~Landsteiner, Y.~Liu and Y.~W.~Sun,
  arXiv:1410.6399 [hep-th].

\bibitem{Liang} 
  T.~Liang, Q.~Gibson, M.~ N.~ Ali,	M.~ Liu, R.~ J.~Cava and N.~ P.~Ong, Nature Materials, {\it to appear}; arXiv:1404.7794 [cond-mat.str-el].

\bibitem{Kharzeev:2014sba} 
  D.~E.~Kharzeev, R.~D.~Pisarski and H.~U.~Yee,
  arXiv:1412.6106 [cond-mat.mes-hall].

\bibitem{Chatterjee:2005mj} 
  S.~Chatterjee, W.~H.~T.~Vlemmings, W.~F.~Brisken, T.~J.~W.~Lazio, J.~M.~Cordes, W.~M.~Goss, S.~E.~Thorsett and E.~B.~Fomalont {\it et al.},
  Astrophys.\ J.\  {\bf 630}, L61 (2005)
  [astro-ph/0509031].

\bibitem{Vilenkin:1980fu} 
  A.~Vilenkin,
  Phys.\ Rev.\ D {\bf 22}, 3080 (1980).

\bibitem{Kusenko:1996sr} 
  A.~Kusenko and G.~Segre,
  Phys.\ Rev.\ Lett.\  {\bf 77}, 4872 (1996)
  [hep-ph/9606428].

\bibitem{Charbonneau:2009ax} 
  J.~Charbonneau and A.~Zhitnitsky,
  JCAP {\bf 1008}, 010 (2010)
  [arXiv:0903.4450 [astro-ph.HE]].

\bibitem{Gorbar:2009bm} 
  E.~V.~Gorbar, V.~A.~Miransky and I.~A.~Shovkovy,
  Phys.\ Rev.\ C {\bf 80}, 032801 (2009)
  [arXiv:0904.2164 [hep-ph]].

\bibitem{Kaminski:2014jda} 
  M.~Kaminski, C.~F.~Uhlemann, M.~Bleicher and J.~Schaffner-Bielich,
  arXiv:1410.3833 [nucl-th].

\bibitem{Shaverin:2014xya} 
  E.~Shaverin and A.~Yarom,
  arXiv:1411.5581 [hep-th].

\bibitem{Redlich:1984md} 
  A.~N.~Redlich and L.~C.~R.~Wijewardhana,
  Phys.\ Rev.\ Lett.\  {\bf 54}, 970 (1985).

\bibitem{Tsokos:1985ny} 
  K.~Tsokos,
  Phys.\ Lett.\ B {\bf 157}, 413 (1985).

\bibitem{Joyce:1997uy} 
  M.~Joyce and M.~E.~Shaposhnikov,
  Phys.\ Rev.\ Lett.\  {\bf 79}, 1193 (1997)
  [astro-ph/9703005].

\bibitem{Frohlich:2000en} 
  J.~Frohlich and B.~Pedrini,
  In *Fokas, A. (ed.) et al.: Mathematical physics 2000* 9-47
  [hep-th/0002195].

\bibitem{Ohnishi:2014uea} 
  A.~Ohnishi and N.~Yamamoto,
  arXiv:1402.4760 [astro-ph.HE].

\bibitem{Grabowska:2014efa} 
  D.~Grabowska, D.~B.~Kaplan and S.~Reddy,
  arXiv:1409.3602 [hep-ph].

\bibitem{Vilenkin:1982pn} 
  A.~Vilenkin and D.~A.~Leahy,
  Astrophys.\ J.\  {\bf 254}, 77 (1982).


\bibitem{Giovannini:1997gp}
  M.~Giovannini and M.~E.~Shaposhnikov,
  Phys.\ Rev.\ Lett.\  {\bf 80}, 22 (1998)
  [arXiv:hep-ph/9708303]; 
  Phys.\ Rev.\  D {\bf 57}, 2186 (1998)
  [arXiv:hep-ph/9710234].


  
\bibitem{Boyarsky:2011uy} 
  A.~Boyarsky, J.~Frohlich and O.~Ruchayskiy,
  Phys.\ Rev.\ Lett.\  {\bf 108}, 031301 (2012)
  [arXiv:1109.3350 [astro-ph.CO]].
  
\bibitem{Boyarsky:2012ex} 
  A.~Boyarsky, O.~Ruchayskiy and M.~Shaposhnikov,
  Phys.\ Rev.\ Lett.\  {\bf 109}, 111602 (2012)
  [arXiv:1204.3604 [hep-ph]].
  
\bibitem{Long:2013tha} 
  A.~J.~Long, E.~Sabancilar and T.~Vachaspati,
  arXiv:1309.2315 [astro-ph.CO].
  
\bibitem{Semikoz:2007ti} 
  V.~B.~Semikoz and J.~W.~F.~Valle,
  JHEP {\bf 0803}, 067 (2008)
  [arXiv:0704.3978 [hep-ph]].
  
\bibitem{Semikoz:2009ye} 
  V.~B.~Semikoz, D.~D.~Sokoloff and J.~W.~F.~Valle,
  Phys.\ Rev.\ D {\bf 80}, 083510 (2009)
  [arXiv:0905.3365 [hep-ph]].
 
\bibitem{Semikoz:2012ka} 
  V.~B.~Semikoz, D.~D.~Sokoloff and J.~W.~F.~Valle,
  JCAP {\bf 1206}, 008 (2012)
  [arXiv:1205.3607 [astro-ph.CO]].
 
\bibitem{Tashiro:2012mf} 
  H.~Tashiro, T.~Vachaspati and A.~Vilenkin,
  Phys.\ Rev.\ D {\bf 86}, 105033 (2012)
  [arXiv:1206.5549 [astro-ph.CO]].
  
\bibitem{Freese:1990rb} 
  K.~Freese, J.~A.~Frieman and A.~V.~Olinto,
  Phys.\ Rev.\ Lett.\  {\bf 65}, 3233 (1990).
  
\bibitem{Kim:2004rp} 
  J.~E.~Kim, H.~P.~Nilles and M.~Peloso,
  JCAP {\bf 0501}, 005 (2005)
  [hep-ph/0409138].
  
\bibitem{Sorbo:2011rz} 
  L.~Sorbo,
  JCAP {\bf 1106}, 003 (2011)
  [arXiv:1101.1525 [astro-ph.CO]].
  
\bibitem{Barnaby:2011vw} 
  N.~Barnaby, R.~Namba and M.~Peloso,
  JCAP {\bf 1104}, 009 (2011)
  [arXiv:1102.4333 [astro-ph.CO]].
  
\bibitem{Alexander:2011hz} 
  S.~Alexander, A.~Marciano and D.~Spergel,
  JCAP {\bf 1304}, 046 (2013)
  [arXiv:1107.0318 [hep-th]].
  
  
\bibitem{Zhitnitsky:2013pna} 
  A.~R.~Zhitnitsky,
  Phys.\ Rev.\ D {\bf 89}, 063529 (2014)
  [arXiv:1310.2258 [hep-th]].
  
\bibitem{Zhitnitsky:2014aja} 
  A.~R.~Zhitnitsky,
  Phys.\ Rev.\ D {\bf 90}, 043504 (2014)
  [arXiv:1404.5965 [hep-ph]].

\bibitem{Ferreira:2014zia} 
  R.~Z.~Ferreira and M.~S.~Sloth,
  arXiv:1409.5799 [hep-ph].

\bibitem{Bartolo:2014hwa} 
  N.~Bartolo, S.~Matarrese, M.~Peloso and M.~Shiraishi,
  arXiv:1411.2521 [astro-ph.CO].
\end{thebibliography}
\end{document}